\documentclass[a4paper,fleqn,usenatbib]{mnras}

\usepackage{newtxtext}
\usepackage{mathptmx}

\usepackage[T1]{fontenc}
\usepackage{ae,aecompl}
\usepackage { float }

\usepackage{graphicx}	
\usepackage{amsmath}	
\usepackage{amssymb}	
\usepackage[section]{placeins}

\usepackage{footnote}
\usepackage{wasysym}

\usepackage[dvipsnames]{xcolor}

\newcommand{\OIII}{[O III]$\lambda$5007}
\newcommand{\OIId}{[O II]$\lambda$7320,7330}

\newcommand{\OI}{[O I]$\lambda$6300}
\newcommand{\NIIt}{[N II]$\lambda$5755}
\newcommand{\NIIr}{[N II]$\lambda$6583}

\newcommand{\NIId}{[N II]$\lambda\lambda$6548,6583}
\newcommand{\SIIb}{[S II]$\lambda$6716}
\newcommand{\SIIr}{[S II]$\lambda$6731}
\newcommand{\SIId}{[S II]$\lambda\lambda$6716,6731}
\newcommand{\SIIIb}{[S III]$\lambda$6312}
\newcommand{\SIIIr}{[S III]$\lambda$9069}
\newcommand{\La}{Ly$\alpha$}
\newcommand{\Ha}{H$\alpha$}
\newcommand{\Hb}{H$\beta$}
\newcommand{\kms}{km s$^{-1}$}
\newcommand{\Mo}{M$_\odot$}

\newcommand{\OIHa}{[O I]$\lambda$6300/H$\alpha$}
\newcommand{\OIIIHa}{[O III]$\lambda$5007/H$\alpha$}
\newcommand{\OIIIHb}{[O III]$\lambda$5007/H$\beta$}
\newcommand{\OIIHb}{[O II]$\lambda$7320+7330/H$\beta$}
\newcommand{\NIIHb}{[N II]$\lambda$6548+6584/H$\beta$}
\newcommand{\SIIHb}{[S II]$\lambda$6717+6731/H$\beta$}

\definecolor{green_5}{rgb}{0.0, 0.5, 0.0}  

\title[Ionized gas properties of Haro 11 with MUSE]{Ionized gas properties of the extreme starburst galaxy Haro 11. Temperature and metal abundance discrepancies.}

\author[V. Menacho et al.]{
V. Menacho$^{1}$\thanks{E-mail: veronica.menacho@astro.su.se}
G. \"Ostlin$^{1}$,
A. Bik,$^{1}$,
A. Adamo$^{1}$,
N. Bergvall$^{2}$,  
L. Della Bruna$^{1}$, \newauthor \space
M. Hayes$^{1}$,  
J. Melinder$^{1}$, 
E. T. Rivera-Thorsen$^{1}$
\\
$^{1}$The Oskar Klein Centre, Department of Astronomy, Stockholm University, AlbaNova, SE-10691 Stockholm, Sweden\\
$^{2}$Department of of Physics and Astronomy, Uppsala University, Box 515, SE-751 20 Uppsala, Sweden
}

\date{Accepted 2021 May 14. Received 2021 April 15; in original form 2020 September 10}

\pubyear{2020}

\hypersetup{draft}

\begin{document}
\label{firstpage}
\pagerange{\pageref{firstpage}--\pageref{lastpage}}
\maketitle

\begin{abstract}

We use high quality VLT/MUSE data to study the kinematics and the ionized gas properties of Haro 11, a well known starburst merger system and the closest confirmed Lyman continuum leaking galaxy. 
We present results from integrated line maps, and from maps in three velocity bins comprising the blueshifted, systemic and redshifted emission. 
The kinematic analysis reveals complex velocities resulting from the interplay of virial motions and momentum feedback.
Star formation happens intensively in three compact knots (knots A, B and C), but  one, knot C, dominates the energy released in supernovae. 
The halo is characterised by low gas density and extinction, but with large temperature variations, coincident with fast shock regions. Moreover, we find large temperature discrepancies in knot C, when using different temperature-sensitive lines. 
The relative impact of the knots in the metal enrichment differs. While knot B is strongly enriching its closest surrounding, knot C is likely the main distributor of metals in the halo. In knot A, part of the metal enriched gas seems to escape through low density channels towards the south.
We compare the metallicities from two methods and find large discrepancies in knot C, a shocked area, and the highly ionized zones, that we partially attribute to the effect of shocks. 
This work shows, that traditional relations developed from averaged measurements or simplified methods, fail to probe the diverse conditions of the gas in extreme environments. We need robust relations that include realistic models where several physical processes are simultaneously at work. 

\end{abstract}

\begin{keywords}
Galaxies: starburst -- Galaxies: ISM -- Galaxy: abundances -- Galaxy: kinematics and dynamics -- Galaxies: star clusters: general -- Galaxies: individual
\end{keywords}


\section{Introduction}

Blue compact galaxies (BCGs) can be considered as local analogues of high-z galaxies due to their  physical properties being similar to those expected for young galaxies in the early universe: they are mostly compact, metal-poor and low-mass galaxies that are experiencing an intense episode of star formation.  Due to their proximity, they provide a close view of the physical process that drives galaxy evolution also at larger distances 
\citep{Kunth2000, ostlin2001, Thuan2005}.

Many BCGs seems to be experiencing some kind of perturbation \citep{ostlin2001},
e.g. gravitational interactions, merging or gas infall, which tend to compress molecular clouds mostly in 
the central region, triggering an intense episode of star formation.
In such a starburst episode, many massive star clusters form 
\citep{Meurer1995,Oestlin1998,Adamo2011}. 
A fraction of these clusters are themselves clustered in compact regions, and can jointly alter the  conditions and dynamics of the gas out to large distances  in the halo \citep{Hopkins2012, Agertz2013, Keller2015}. 

Massive star clusters contain thousands of massive stars that continuously  inject  large amounts of energy, heat and momentum to their surrounding gas \citep{Kim2015,Kim2017, Agertz2013}. At younger ages, massive stars produce an intense ionizing radiation field whose energetic photons are able to reach far out into the halo. At the same time, powerful stellar winds and radiative pressure in dusty environments inject momentum that produces turbulence and creates a porous medium facilitating in this way, the transport of both ionizing photons and matter out of the star forming region \citep{Silk1997, Weilbacher2018}. Later on, this momentum injection is considerably enhanced by supernova explosions, that in case of being recurrent, can alter the condition of the gas at large scales. 
This stellar feedback can be powerful enough to create multi-scale structures such as filaments, arcs, shells, bubbles, cavities and rings \citep{Tenorio-Tagle1988,Hunter1993,Marlowe1995,Hopkins2012,menacho2019} and to drive outflows on scales up to kiloparsecs  \citep{Tenorio-Tagle2006,Heckman2017,Fielding2018,Bik2018}. 
Moreover, given the shallow gravitational potential in low mass galaxies, strong feedback is able to transport metal enriched gas further out in the halo, or even more, to launch galactic winds carrying metal-enriched gas to the intergalactic medium (IGM) and clearing paths through which ionizing Lyman continuum (LyC) photons can escape \citep{Heckman2015,Chisholm2015}. 

The mechanisms that favour the escape of LyC photons are not well understood. For many years only a handful of galaxies were found to leak LyC radiation, however this number is rapidly rising thanks to campaigns targeting galaxies with highly ionized gas (high [O III]/[O II] ratio) or close \La\ peak separation  \citep{Verhamme2017,Izotov2018,Izotov2018a, Jaskot2019}, quantities which prove to be successful indirect tracers of LyC leaking galaxies. Despite that, very little is known about the physical properties of their gas and the processes that govern the escape of LyC radiation, mostly because most of these galaxies are compact and/or distant (high-z). 

This paper focuses on Haro 11, the first and nearest confirmed LyC leaking  galaxy \citep{bergvall2006, leitet2011}. Haro 11 is a merger system and one of the most extreme starbursts in the local universe \citep{ostlin2015}. 
In the present starburst, that peaked about 3.5 Myrs ago, at least two hundred massive star cluster with masses ranging from 10$^4$ to 10$^{7}$ \Mo were formed. Most of these clusters are found in three condensed regions: knot A, B and C in the central starburst \citep{Adamo2010}. 
The large number of young massive stars produce an intense ultraviolet radiation field, which is reflected in the amount of ionized gas, whose mass is about three times greater than the neutral atomic gas mass \citep{MacHattie2014,pardy2016, menacho2019}. 

In a previous paper \citep{menacho2019} we sliced up the key emission lines into several velocity bins, and studied the impact of strong stellar feedback on the interstellar medium (ISM) of the galaxy. 
Using \Ha\ and \OIIIHa\ maps as tracers of the ionized gas architecture and the degree of ionization in the gas, we showed that Haro 11 has developed kpc-scale filaments, arcs, shells, tidal tails, ionized cones and outflows, appearing in specific velocity ranges. 
We briefly describe some of these structures, that are relevant for this work, in section \ref{subsection:structuresIonGas}. 
We also found that about one fourth of the total ionized gas mass will likely escape the galaxy. 
Thus, Haro 11 provides a unique opportunity to study a) the impact that strong radiation and momentum feedback has on the ISM of a galaxy, b) the interplay between merger dynamics and stellar feedback and c) the mechanisms that favour LyC leakage.

In this work we focus on the kinematics and the physical properties (density, temperature, dust extinction and metallicity) of the ionized gas. 
These properties are analysed from integrated line maps and from maps separated into three velocity bins tracing the gas at blueshifted, central and redshifted velocities.
The maps in velocity bins are used to investigate the conditions of the ionized ISM in the structures  previously identified.

This paper is organised as follows: in Section 2 we summarise the observations, data reduction and methods applied in our analysis. In Section 3 we derive the systemic velocity of Haro 11 and briefly describe the main ionized gas structures that defines the halo. In Section 4 we present an analysis of the kinematics derived from the recombination \Ha\ line and the metal-lines \NIIr\ and \OIII . In Section 5 we analyse the density, temperature and extinction properties of the gas. In Section 6 we derive the oxygen, sulphur and nitrogen abundances following a temperature-dependent method. We then compare our results with  metallicities derived from a strong-line method. In Section 7 we discuss the metallicity discrepancies found in both methods and additionally analyse the production places of metals in Haro 11. Finally, we present a summary and conclusions in Section 8.

\section{Observations, data reduction and methods}

Haro 11 was observed at VLT with the Multi-Unit Spectroscopic Explorer MUSE in the Wide-field mode \citep[MUSE]{bacon2010} 
in several nights between December 2014 and August 2016. The observing strategy, data reduction and methods are presented in \citet{menacho2019} in detail, which we here briefly summarise. 
The galaxy was observed in the extended wavelength setting in 4 pointings arranged in a 2x2 mosaic (contiguous pointing overlap by 30\arcsec ). 
For each pointing eight exposures of 700s were obtained.
The total exposure depth of the final 1.5\arcmin x1.5\arcmin\ mosaic thus varies with the deepest data in the central 30\arcsec x30\arcsec\ (with a total integration time of $>6$h)  and the lowest in the corners ($\sim 1.5$h).
 
The data was reduced with the MUSE pipeline  \citep[version 1.2]{weilbacher2012}, following the standard reduction procedures. 
The final cube  covers a wavelength from 4600 to 9350 {\AA}. The wide-field mode has a spatial sampling of 0\arcsec.2 x 0\arcsec.2, a spectral element of 1.25 {\AA} and a spectral resolution of $\sim$100 \kms\ for \Ha\ at the redshift of Haro 11. The image quality of the final cube is about 0.8\arcsec .
We removed the underlying stellar absorption affecting the Balmer emission lines by running the python package \texttt{pPXF} \citep{Cappellari2004,Cappellari2017}.
A difference of the \citet{menacho2019} analysis, we perform a new \texttt{pPXF} run with the improved stellar spectra library \texttt{E-MILES} \citep{vazdekis2016}. This steps insured a better removal of the underlying stellar absorption

\subsection{Extracting line maps}\label{subsec:extractingMaps}

We produced two types of maps: integrated line maps and velocity channel maps. The former were obtained by integrating the continuum-subtracted emission from the entire line spectra, while in the latter we integrated the emission only from parts of the line spectra within specific velocity bins. 
These line maps  were created from the final science cube, and 
from the stellar absorption-corrected science cube for the lines affected by stellar absorption (\Ha , \Hb , and for the \SIIIb\ line that is affected by a close stellar absorption feature).
For each of those line map we extracted their associated error map from the square root of the final variance cube.

Before extracting maps in velocity channel, we have to ensure that all lines involved in the estimations of each of the ISM properties need to have the same spectral resolution. This is important because in MUSE spectral lines are broadened as function of their wavelengths (i.e. the spectral resolution is wavelength dependent). In our data, the spectral resolution given by the Full Width at Half Maximum (FWHM) of the MUSE Line Spread Function (LSF) ranges from 2.45 to 2.9 \AA , as calculated from the equations 7-8 from \citet{Bacon2017}. Thus, for each diagnostic (i.e. temperature, densities, abundances, reddening) we standardized the spectral resolution of the lines involved to the lowest resolution of these lines.
We do so by convolving the line spectra with a 1D Gaussian using the task \texttt{fftconvolve\_gauss} from the Muse Python Data Analysis Framework (\texttt{MPDAF}) \citep{Bacon2016}.
These spectra were then resampled using spline interpolation to 50 \kms\ per spaxel. 

We selected three velocity channels based on a visual inspection of a sequence of \Ha\ maps in 50 \kms\ bins, over a velocity range of -400 to 350 \kms , presented in \citet{menacho2019}. 
We notice that repeating pattern of specific Ha structures predominate in the halo either at blueshifted (i.e. southern filaments, bright \Ha\ arc), redshifted (i.e. longest S-E tidal tail, dusty arm) and central velocities (loop-like tidal tails). These \Ha\ structures are seeing in Fig. \ref{fig:HaFlux} and further described in section \ref{subsection:structuresIonGas}.
Thus, we extracted integrated line maps from the resampled spectra in the following velocity channels: 
from -325 to -75 \kms\ (channel width$=$250 \kms ) tracing the approaching (blueshifted) gas;
from -75 to 75 \kms\ (width$=$150 \kms ) for the gas at central velocities and from 75 to 325 \kms\ (width$=$250 \kms ) for the receding (redshifted) gas.
Using a broad velocity width for the blue- and redshifted channels is problematic when disentangling two nearby lines (e.g. \SIId). Additionally, gas components of both galaxy progenitors are superimposed along the line of sight in some areas, which broadens the lines even more. 
In those cases, we integrated the flux of the line in question down to the wavelength where both lines intersect. 

Once the integrated and channel maps were extracted, we subsequently normalised, for each diagnostic, the spatial resolution of the lines that are considerably separated in wavelength. 
Sky residuals (emission or absorption) were corrected in all maps by subtracting the median value from a customised sky region, that was created by Voronoi-binning the map (signal-to-noise threshold S/N $=$ 3 and maximal area of 1.8 \arcsec ) and masking the regions with SNR S/N $\geq$ 3. 
The \SIIb , \SIIr\ and \SIIIr\ lines fall in a spectral range affected by telluric absorption. Although these features were corrected during the data reduction processes, some spikes still persisted in the final cube. The contribution from these spikes are included in the sky residuals and hence removed from the maps with the sky residuals. 
For each diagnostic, the maps involved where then adaptively binned using Voronoi Tesselations \citep{cappellari2003,diehl2006} with a minimum signal to noise value ($S/N \ge $5) over the weakest line, and a maximal area based on the physical parameter to derive. 

The fluxes from the line maps that are used for the temperature and metallicity estimations were subsequently corrected for dust attenuation. For this purpose, we applied the Voronoi pattern of the diagnostic to the \Ha\ and \Hb\ line maps, and computed the nebular reddening E(B-V) using the Balmer decrement and the Cardelli attenuation curve \citep{Cardelli1989, Osterbrock2006}. The fluxes were then corrected for the estimated nebular reddening and the foreground Milky Way reddening (E(B-V)$=$0.01 mag at the position of Haro 11  \citep[NED]{Schlafly2011}, while simultaneously the uncertainties from the error maps were propagated accordingly.

These line and error maps from the integrated and channel maps were finally used for the estimations of the ISM properties.

\subsection{Extracting aperture quantities from the galaxy and the star forming knots}\label{subsec:extractingSpectra}

We extracted an integrated spectrum from the entire galaxy using a 1\arcmin\ aperture (about 26 kpc); along with integrated spectra from 1\arcsec\ apertures (diameter $\sim$ 0.43 kpc) centred in the three knots A, B and C, and a star forming region in the dusty arm (see Fig. \ref{fig:HaFlux}). All these apertures are centred on the \Ha\ emission peak from their respective regions. 
These aperture spectra are used to compute the recessional velocity of the Haro 11, which described is in section \ref{section:systemicVel}.

Next, we derive integrated flux-weighted ISM properties from the aforementioned apertures, that provide further information to the spatially resolved maps.
For this, we compute integrated fluxes and errors over the apertures from the previously created line and error maps from both, the integrated and channel maps (see section \ref{subsec:extractingMaps}). We then compute the E(B-V) values in the given apertures, and corrected for dust attenuation (using the Cardelli attenuation curve) the fluxes from the lines involved in the temperature and metallicity estimations. Finally, we derive the E(B-V), density, temperature, and metallicities following the recipes described in the sections 5 and 6. We note that in each operation on the integrated fluxes, the errors were propagated accordingly.

\subsection{Error estimations}

The errors from the apertures reported in table \ref{tab:param_knots}, and the ones used in the S/N calculations in the maps, are estimated as follows. 
For the kinematics we report the errors from the fitting procedures. In the reddening and in all abundance estimations the uncertainties from the input parameters were propagated accordingly each time we performed an operation on the lines involved. Finally, for the density and temperatures we estimated the errors using the Monte Carlo technique in the following manner. For each measurement (Voronoi cell or  aperture) we generated 1000 realisations of the line fluxes involved in the respective diagnostics.  Each of these realisations followed a Gaussian distribution, centred on the line flux, with $\sigma$ equal to the 1$\sigma$ uncertainty associated with the line flux. This procedure provides a distribution of 1000 possible density and temperature estimations for each Voronoi cell or aperture measurement. Finally, we reported the upper and lower errors from the uncertainties taken from the 16\% and 84\% confidence intervals. We note however that for knot C we were unable to estimate the errors of the densities from the blueshifted and redshifted gas, because the sampled values were lower than the minimum value needed to compute this property from the [SII]-line diagnostic. This applies also for the missing lower errors in the density and [NII]-temperatures of knot A and/or C, as shown in table \ref{tab:param_knots}.

\section{The systemic velocity and the ionized gas components of Haro 11}\label{section:systemicVel}

\begin{figure*}
    \centering
    \includegraphics[width=2\columnwidth]{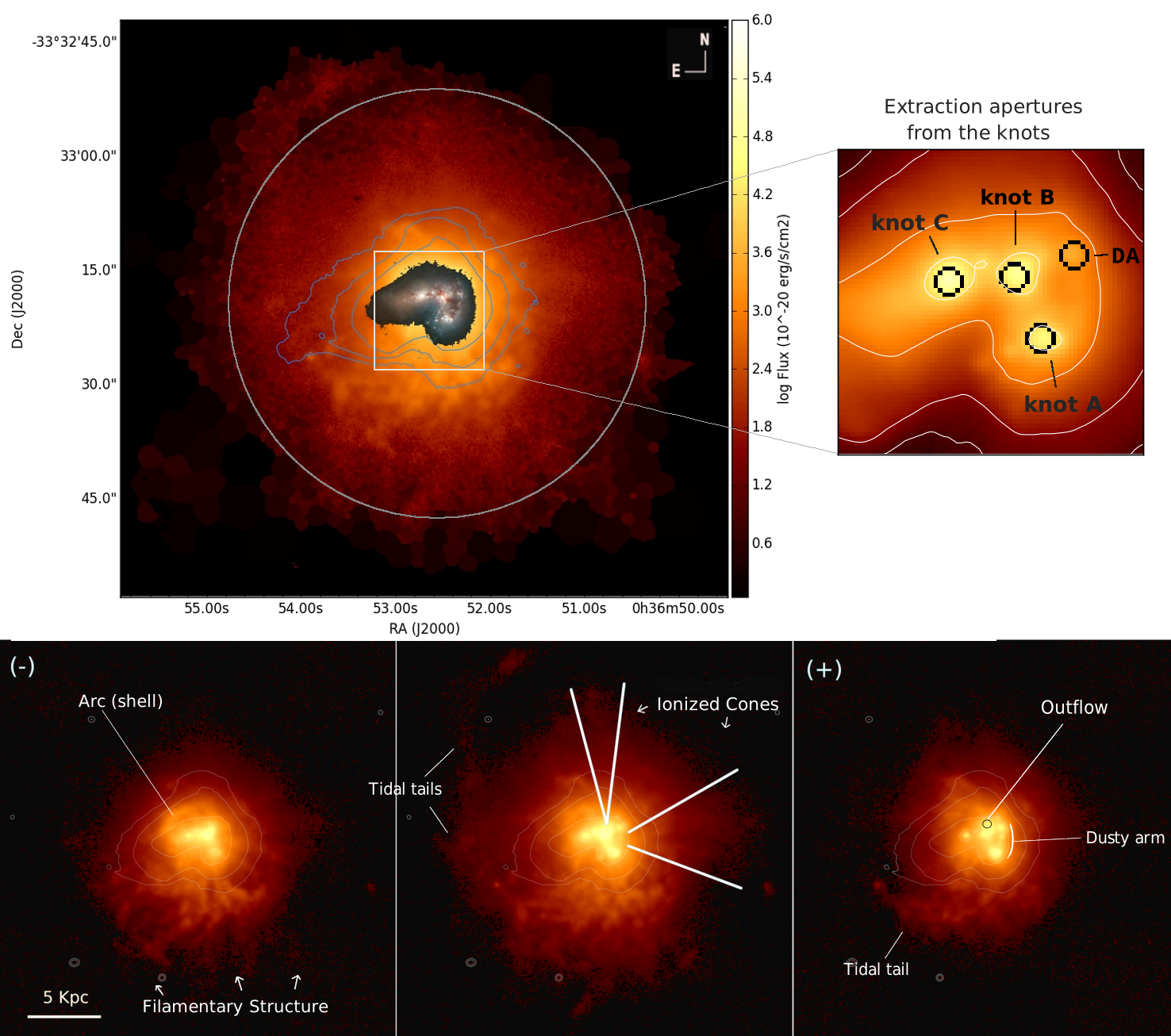}
    \caption{Top left: Integrated H$\alpha$ map in its full extent with the HST composite image superimposed in the centre. Here and in all maps, the overlaid contours show the MUSE I-Band stellar continuum to provide a spatial reference in the maps. The white circle shows the 1\arcmin\ aperture in diameter covering the entire galaxy. of the  Top right: Continuum image showing the zoom in view of the central region highlighted in the left panel. In black we show the extraction apertures of the knots A, B and C, and the star forming knot in the dusty arm (DA). 
    Second row, from left to right: \Ha\ maps of the blueshifted (-), central and redshifted (+) \Ha\ gas showing the main features at these specific velocity bins. These are: the filamentary complex and the bright arc at blueshifted and central velocities; the large tidal tails product of the merger dynamics at central and redshifted velocities; and the bright outflow and the dusty arm at redshifted velocities. We additionally trace in white lines the position of the ionized cones that were identified in the \OIIIHa\ map (see Fig.3 in \citet{menacho2019}) at central velocities. HST image credit: ESO/ESA/Hubble and NASA. 
    }
    \label{fig:HaFlux}
\end{figure*}

Fig. \ref{fig:HaFlux} shows the integrated Voronoi-binned (S/N$\geq$5) H$\alpha$ map with the HST composite image of Haro 11 in the centre. 
At a sensitivity of 1.5 $\times$ 10$^{-20}$ erg s$^{-1}$ cm$^{-2}$ at 5$\sigma$, the extended \Ha\ halo is about 7 times larger in radial direction than the 4$\times$4 kpc$^2$ central region where star formation mostly happens. 
Despite that Haro 11 is a merger complex with the main body similar to the Antennae galaxy \citep{ostlin2015}, the diffuse halo-gas shows a circular symmetry at all velocities. 
Such large halos are commonly seen in starburst dwarf galaxies and constitutes an evidence for the presence of an intense radiation field of energetic photons, produced by a numerous population of young and massive stars \citep{Bergvall2002, Ostlin2014, Bik2018}.

\begin{table}
\resizebox{1\columnwidth}{!}{%
    \begin{tabular}{lcccc}
    \hline
    Line      & $\lambda_0$ & $\lambda_{obs}$       & v$_{res}$       & ${\Delta}vel$   \\ 
          &  [\AA] & [\AA]       & [\kms ]         &  [\kms ]  \\ \hline
    Ha                    & 6562.8    & 6698.34  $\pm$0.06 & 6195.83  $\pm$2.82 & 1.83  $\pm$5.6   \\
    Hb                    & 4861.3    & 4961.64  $\pm$0.01 & 6192.17  $\pm$0.35 & -1.83 $\pm$10.5  \\
    $[$OIII$]\lambda$5007 & 5006.8    & 5110.36  $\pm$0.08 & 6205.16  $\pm$4.54 & 11.16 $\pm$6.7 \\
    $[$NII$]\lambda$6583  & 6583.5    & 6719.04  $\pm$0.03 & 6176.35  $\pm$1.31 & -17.65$\pm$5.1 \\
    $[$SII$]\lambda$6716  & 6716.4    & 6854.93  $\pm$0.21  & 6187.69  $\pm$9.25 & -6.31 $\pm$10.5 \\
    $[$OI$]\lambda$6300   & 6300.0    & 6430.29  $\pm$0.10 & 6204.29  $\pm$4.69 & 10.29 $\pm$6.8 \\ \hline
    \end{tabular}}
    \caption{Best fit of six line spectra taken from the integrated flux in an aperture of 1\arcmin ($\sim$26 kpc). Column 1-3 shows the list of the emission lines, their rest-frame wavelength and the peak of the observed lines in wavelength respectively. Column 4 shows the calculated recessional velocities from each line. 
    The galaxy recessional velocity of 6194$\pm$4.9 \kms , corresponding to the redshift of 0.0206467, was calculated from the median velocity of the lines. In the fifth column (${\Delta}vel$) we show the velocity difference of the different lines with respect to the central velocity of the galaxy.}
    \label{tab:fit_z}
\end{table}

Owing to the high sensitivity gained in our data and the large field of view of our observations, we compute the systemic velocity of Haro 11 from the integrated 1\arcmin\ aperture spectrum, by performing  simple Gaussian fits to the six strongest lines used in this paper. The fit of some of these lines are seen in fig. \ref{fig:fit_spectra_gal}) in the appendix. The results are presented in Table \ref{tab:fit_z}. The median galaxy recessional velocity of 6194 $\pm$ 4.9 \kms\ is calculated from the peaks of the Gaussian fits, and the error is propagated from the uncertainties reported from the fitting procedures. This recession velocity, which is used in the remaining of this paper, is 15 \kms\ larger than the recession velocity used in \citet{menacho2019}.
From this value we calculate the redshift of Haro 11 of z$=$0.0206467. 
We observe that the highly and lowly ionized gas (traced by the \OIII\ and  \OI\ respectively) is slightly redshifted and the intermediately ionized gas (\NIIr\ and \SIIb\ lines) blueshifted, but within the errors, relative to the gas that recombines (traced by the \Ha\ and \Hb\ lines).

\subsection{Summary of the ionized gas structures in Haro 11}\label{subsection:structuresIonGas}

The ionized gas architecture of Haro 11 was already analysed in \citet{menacho2019}, where several structures were uncovered from maps of the \Ha\ and the ionization mapping sliced in velocity bins of 50 \kms.  In this work we analyse the physical properties of some of these components labelled in the \Ha\ maps of Fig. \ref{fig:HaFlux}. In the lower panels of Fig. \ref{fig:HaFlux} we show from left to right maps of the blueshifted, central and redshifted gas. Structures such as filaments, arcs, bright clumps and tidal tails become clearly visible outside the star forming regions. Here we briefly describe them in order to orient the reader in the next sections.

\begin{itemize}
    \item {The \textit{filamentary structure} comprises all the radially oriented filaments that cover the southern hemisphere at blueshifted and central velocities. This structure is highly ionized. \citet{menacho2019} suggested that they were formed from a kpc-scale superbubble breakout, whose vented hot gas dragged away dense gas clumps and part of the shell forming the filaments. This large superbubble was likely developed by supernovae from all star clusters in the central starburst.} 
    \item {The bright \textit{arc} encloses the starburst region towards the north and is visible at blueshifted and central velocities. This arc might be shell-remnant of the superbubble that has developed the filamentary structure in the southern part.}
    \item {A shell of a second, smaller (r $\sim$ 1.7 kpc), somehow fragmented \textit{superbubble surrounding knot C} was evidenced in the ionization mapping at high blueshifted and redshifted velocities. This structure, which is not shown in Fig. \ref{fig:HaFlux}, seems to lean on the arc structure towards the north-east of knot C.}
    \item {Two galactic-scale \textit{ionized cones} (white lines in Fig. \ref{fig:HaFlux}) were uncovered in the ionization mapping (\OIIIHa ) at central velocities. The northern one originate at the bases of knot B, while the western one was likely developed jointly by knots A and B.}
    \item {A bright \textit{outflow} just in the northern part of knot B, but redshifted, was traced with velocities up to 1000 \kms\ in the \OIII\ line. This outflow stands out in all emission lines (see \textit{2} in the $\sigma$ map in Fig. \ref{fig:velocity}).}
    \item {Three tidal tails product of the merger dynamics are traced farther out in the halo at central and somewhat redshifted velocities. Two of them form a loop along the line-of-sight, and one, that is also visible at redshifted velocities, stretches out to the east. Because these tails are faint, we were not able to determine the physical properties of their gas, except for one that is highlighted at redshifted velocities whose gas properties are intrinsically in the properties of the gas at redshifted velocities. 
    }
\end{itemize}{}

\section{Kinematics of Haro 11}\label{sec:kinematics}

\begin{figure*}
    \centering
    \includegraphics[width=17.8cm]{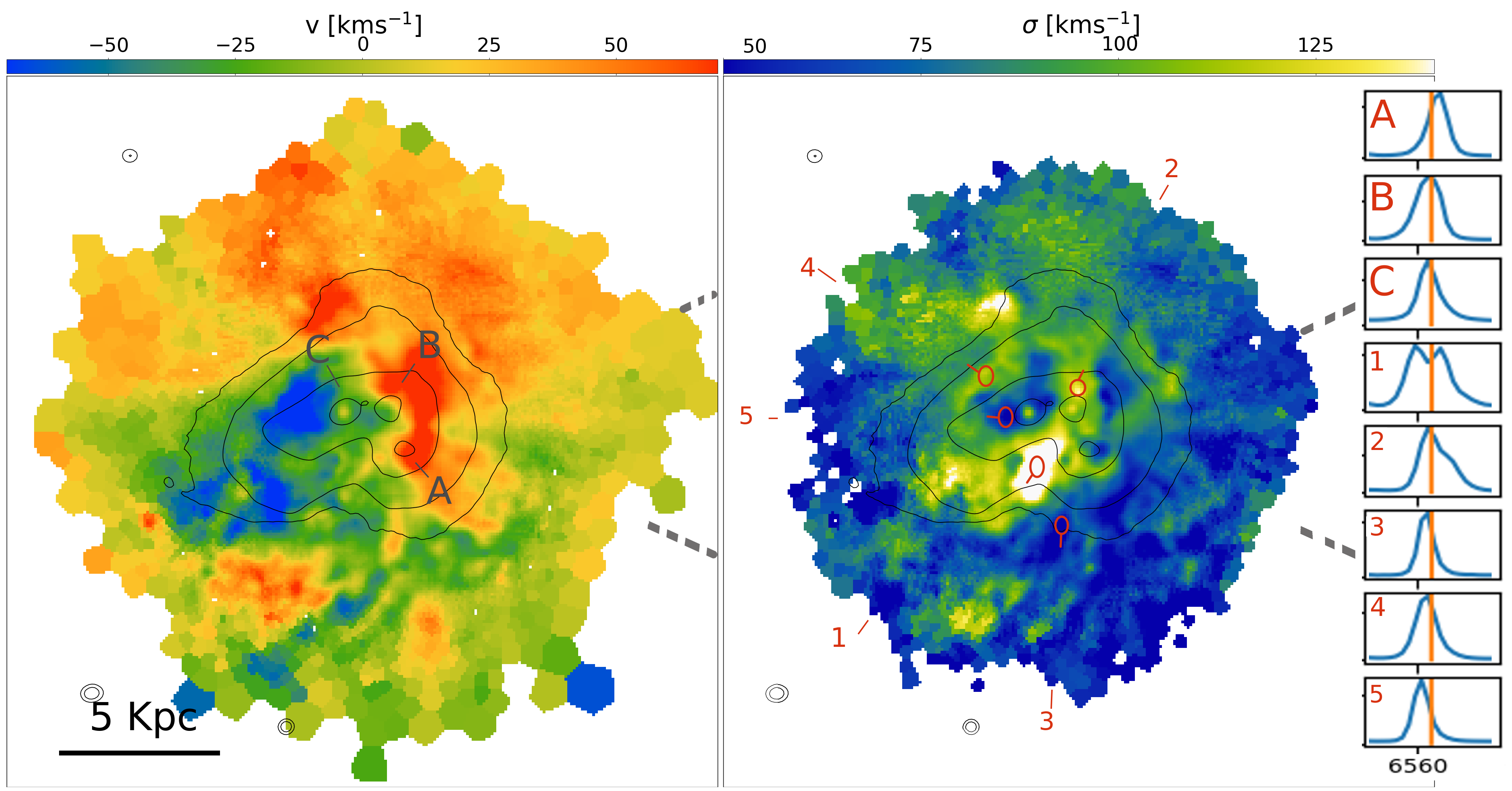}
    \caption{\Ha\ Kinematics of Haro 11. The left and right maps are the velocity and velocity dispersion derived from a single Gaussian component fit. The velocity ranges from -110 to 110  \kms\  and the velocity dispersion from 30 to 170 \kms . In black we show the stellar continuum and in dashed lines the projected location of the western ionized cone. 
    On the $\sigma$ map we show in red circles five relevant regions labelled from 1-5. 
    The inset panel on the right end side shows the \Ha\ profiles from the 1\arcsec\ aperture spectra of the knots A, B and C; and from a 0.6\arcsec\ aperture spectra of the five regions of interest that were extracted from the centre of the red circles. The orange vertical line is the position of the \Ha\ at systemic velocities. The numbered spectra correspond to the following regions of interest: 
    (1) highest velocity dispersion, (2) spectra from the position of an outflow at the bases of knot B, (3) low velocity dispersion region coincident with a filament seen at blueshifted velocities, (4) the arc structure, (5) the low velocity dispersion east of knot C.
    }
    \label{fig:velocity}
\end{figure*}

The kinematics are derived from a single Gaussian fit to the continuum-substracted \Ha\ line spectra of each pixel and we use the first and second moment to derive the velocity and FWHM information.
The velocity dispersion was then derived from the FWHM ($\sigma= \text{FWHM}/2.355$) after the subtraction of the instrumental broadening \citep{Bacon2017}. 
With the aim to extract the information to the farthermost part of the halo, we create a Voronoi pattern of the integrated \Ha\ emission with S/N $\geq$ 40 and a maximal cell size of 7.7 arcsec$^2$ and applied this to the velocity and velocity dispersion maps\footnote{We verified that applying the Voronoi pattern to our velocity maps do not change the the overall kinematic pattern in the galaxy.}. 

Despite that Haro Haro 11 is an ongoing merger with multiple kinematic components, we use a single-component fit to map the mean velocity, weighted by the flux, from the emitting gas; and to map the regions with narrow and broad line profiles. The velocity information is then the flux-weighted mean velocity of all possible gas components along the line of sight. Because of the weighting, the mean velocity is biased towards the component with the strongest emission. 

The left panel of Fig. \ref{fig:velocity} shows the complex \Ha\ kinematics up to a distance of 13.5 kpc in radius. 
The velocity ranges from -110 to 110 \kms\ from the east of knot C to the west of knot B respectively.
The blueshifted gas does not reach our detected \textit{edges} of the halo, but seems to be connected to knot C. This gas likely originates from one of the two progenitors. 
On the other hand, the redshifted gas widely extended towards the northern halo, the dusty arm, knots A and B may originate in the other gas rich progenitor. 

\citet{James2013} and \citet{ostlin2015} have suggested rotation velocities based on the velocity gradient seen in the central part (r $<$ 4 kpc). 
Although there are signs of a global rotation at a PA $\sim$ 35 deg in the outer halo (r $>$ 4 kpc), the global motion of the gas appears to be setup by a mixture between virial motions and strong stellar feedback.  Haro 11 is a late-stage merger and as such, the outer halo is expected to have reacted to the gravitational potential earlier in the interaction and to have started rotating. 
On the other side, the starburst condition in Haro 11 have produced an intense stellar feedback which rapidly developed multi-scale structures (e.g. filaments, arcs, bubbles) and outflows, that have partially washed away the imprints of the merger dynamics.  

Imprints of virial motions are thus the velocity gradient in the central part, which seems to show the motions of both galaxy progenitor cores, knots B and C, in process of coalescence; and the tidal tails (see Fig. \ref{fig:HaFlux}) originated by gravitational torques from the interaction of the galaxy progenitors. 
Imprints of stellar feedback, on the other side, dominate the south-western hemisphere. 
In the southern halo, the core of the line is mostly blueshifted by the bright filamentary structure, that, as suggested in \citet{menacho2019}, results from a kpc-scale superbubble breakout. 
Besides, in the position of the western ionized cone, projected by grey dashed lines in Fig. \ref{fig:velocity}, the gas is slightly blueshifted compared to its surrounding gas which can be attributed to the effect of feedback. 

Haro 11 was dubbed as a miniature-Antennae by \citet{ostlin2015} as it morphologically resembles the Antennae galaxy. The kinematics of both galaxies are, however, somehow different. While merger-induced virial motions dominate the kinematics of the Antennae galaxy \citep{Weilbacher2018}, we find that in Haro 11 both, virial motions and strong stellar feedback from the vast population of massive star clusters are equally affecting the global kinematics of the galaxy. 

In the knots, the velocities suggest that knots B and C are sitting in the kinematic centre of the galaxy,
while knot A and the dusty arm are more redshifted (see table \ref{tab:param_knots}). The flux weighted mean recessional velocities of the knots are shown in table \ref{tab:fit_vel_knots} in the appendix.
Since knot A\footnote{Knot A is the region south of knot B that is bright in line emission and continuum. This area, which is more extended than knots B and C, contains several massive star clusters that are well separated in the HST image \citep{Adamo2010}} and the southern part of the dusty arm coincide along the line-of-sight, we carefully inspected the velocities and line shapes in the extended area of knot A, 
aiming to spatially disentangle both components. 
We clearly identified two components in this knot separated by ${\Delta}v =$ 40 \kms . The western part (western edge in knot's A contour), which contains some star clusters \citep{Adamo2010}, has redshifted centroid velocities compared to the remaining area of knot A (about 75\%) and seems to trace a gas that belongs to the dusty arm.
Comparing the velocities of the ionized gas and the stars (provided by \texttt{ppxf}) in the knots, we found that both components agree to some extent well (${\Delta}$v$_{IonGas-Stars} \sim$ 20 $\pm$ 25 \kms ). The same result was found by \citet{ostlin2015} who suggest that both components are in general terms set by virial motions. 

The right panel in Fig. \ref{fig:velocity} shows the velocity dispersion map up to a distance of 10.5 kpc in radius. 
We measure an average velocity dispersion of 116 $\pm$ 3 \kms\ for the entire galaxy and an intrinsic velocity dispersion ($\sigma_{0} = \sum (Flux^{H\alpha}_{pixel} \times
 \sigma_{pixel}) / \sum (Flux^{H\alpha}_{pixel}) $) of 91 \kms , which is similar to the value obtained by \citet{ostlin2015} within the errors. 
The latter is the flux-weighted average velocity dispersion and measures the average random motion of the ionized gas, while the observed $\sigma$ includes the effect of rotations and other velocity amplitudes. 

The line profiles in most parts of the galaxy do not resemble a perfect Gaussian shape, especially in the broadest line region. In some cases the line shape shows hints of multiple gas components along the line of sight,  while in other cases this is not obvious. Nevertheless, in most broad line regions the high line widths might result from two or more resolved or unresolved narrower components.

On the $\sigma$-map we show five regions of interest marked with red circles and labelled from \textit{1-5} that we analyse next. The right inset on the sigma map shows the \Ha\ profile from the knots and from the regions of interest, extracted from an integrated 0.6\arcsec\ aperture diameter within the red circles.
The velocity dispersion ranges from 30 to 170 \kms\ with a flux-weighted median of 103$\pm$3 \kms\. 
Low values ($\leq$50 \kms ) are measured in the well defined filamentary structure (e.g. region \textit{3}) and in the somewhat intriguing elongated cavity surrounding knot C (region \textit{5}). 
On the other hand, larger dispersion velocities dominate the central region and the northern redshifted halo. In the regions \textit{1, 2} and \textit{4} we identify some discrete structures characterised by local maxima. In region \textit{1} we show the broader line region whose line profile clearly shows a double kinematic component, suggesting that there, the gas of both galaxy progenitors overlap along the line of sight. In region \textit{2} we show the position of the bright redshifted outflow, traced up to 1000 \kms , and in region \textit{4} we clearly identify the bright arc seen in the \Ha\ maps.

At large scales it becomes evident that the highest velocity dispersion values are measured mostly in the gas surrounding knot C at r$>$ 1.5 kpc.
Supernova-driven blastwaves and stellar winds from the massive stellar population in knot C, are likely inducing turbulence not only in the arc (shell) structure but in most of the north-east halo. 
Knot C is likely the nuclear star cluster of one galaxy progenitor and we will show later that it dominates by far the energy released by supernova in the galaxy (section \ref{sec:SN_knots}). This knot seems to be sitting within a shell of low density gas. In section \ref{"subsec:density"} we estimate densities lower than 1 cm$^{-3}$ in the cavity surrounding knot C (region \textit{5}, blue colour in $\sigma$ map) in the integrated and channel maps.
Thus it is possible that shocks triggered by supernovae propagate unhindered outwards, impacting directly the arc structure or even in the more distant gas structures in the halo. 
The western ionized cone is other structure that shows enhanced velocity dispersion (${\Delta}_\sigma \sim 25$\kms), specially in its foundation (grey dashed line in Fig. \ref{fig:velocity}). 
The radial stripes seeing in the map suggest an outflowing gas, that produce turbulence by shocking the halo gas in its way out. The kinematic information here hint at a galactic outflow of highly ionized gas. 
Evidence of shocks from 200 to 300 \kms\ impacting on the eastern-hemisphere of Haro 11 are presented in a future paper (Menacho et al, in prep.) using the [OI]-BPT diagram and shock models of \citet{Allen2008}. In the western ionized cone we find a combination of photoionization and shocks as the main excitation mechanisms of the gas.

Comparing the velocity dispersion of the gas and stars in the knots, we find the gas component to be kinematically hotter. \citet{ostlin2015} on the other side found similar $\sigma$ for the gas and stars. These authors seem to trace the gas from HII-regions in the knots that is mostly governed by the merger dynamics, while we might be tracing also the gas that is affected by feedback (outflows, shocks).

\subsection{Kinematics from the \NIIr\ and \OIII\ metal lines}\label{subsec:kinematics}

\begin{figure}
    \centering
    \includegraphics[width=1\columnwidth]{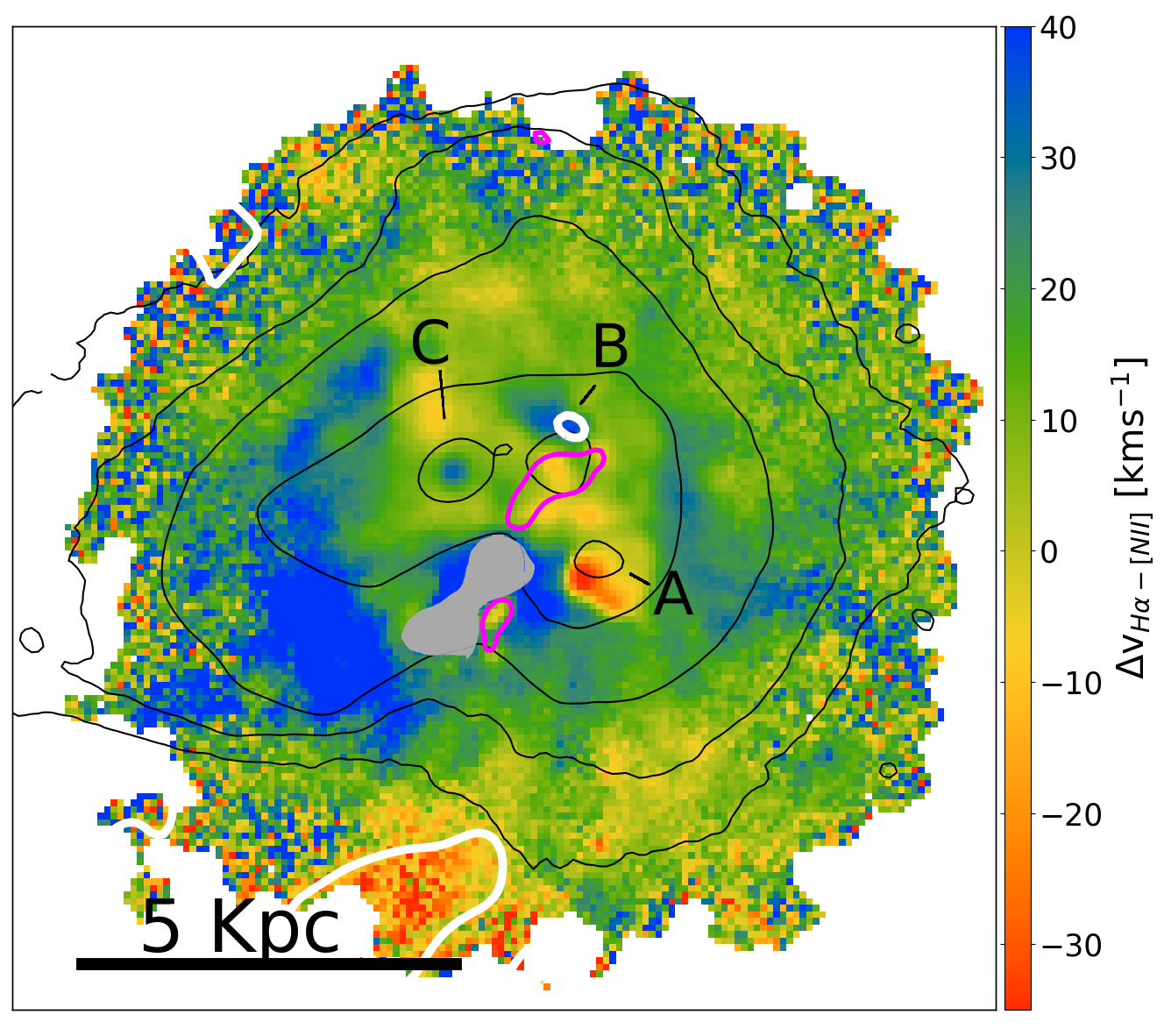}
    \caption{Velocity difference of the \Ha\ and [N II] emitting gas. Positive (negative) values indicate blueshifted (redshifted) [N II]-velocities compared to the \Ha\ velocities.  The broad line region is masked in grey (filled area) to avoid unreliable values due to the possible contamination of the broad \Ha\ line in the [N II] line.
    The velocity difference ranges from -40 to 70\kms , with the [N II] gas being predominantly blueshifted.
    In magenta we delineate the contours of two strongly [N II]-enriched areas (see Fig. \ref{fig:element_abundance}), that are associated with Wolf-Rayet stars. 
    Additionally, we show in white contours the regions where the \OIII\ gas is blueshifted by more than 20 \kms\ compared to the \Ha\ velocities. This condition is fulfilled in part of the filamentary structure and in the outflow close to knot B.
    }
    \label{fig:velocityHaNII}
\end{figure}

In addition to the \Ha\ kinematics, we derived the [N II] and [O III] kinematics from the \NIIr\ and \OIII\ lines.
These lines are tracing a slightly different ambient gas in terms of ionization, namely a moderately and highly ionized warm gas respectively. 
Furthermore, the strength of their lines depend on the gas density and the respective nitrogen or oxygen-abundance from the gas they originate in.
Nevertheless, we will broadly compare their velocities with respect to the \Ha\ velocities presented previously.

\begin{table*}
\normalsize
\centering
    \begin{tabular}{lllccccc}
    \hline
    ~                    & ~              & ~ & Galaxy integrated  & Knot A           & Knot B            & Knot C             & Dusty arm \\ 
    \hline
    vel$_{H\alpha}$     &[\kms ]        & ~  & 2  $\pm$3         & 60 $\pm$2         & -22 $\pm$4        & -7 $\pm$3         & 98 $\pm$2       \\
    vel$_{[N II]\lambda6583}$ &[\kms ]    & ~ &-18 $\pm$1         & 75 $\pm$2         & -19 $\pm$2        & -32 $\pm$1        & 86 $\pm$2        \\
    vel$_{[O III]\lambda5007}$ &[\kms ]   & ~ & 11 $\pm$4         & 67 $\pm$4         & -6 $\pm$4         & -31 $\pm$3        & 104 $\pm$3       \\
    $\sigma_{H\alpha}$   &[\kms ]        & ~ & 116 (90*) $\pm$3   & 84$\pm$3          & 106$\pm$5         & 89$\pm$5          & 70$\pm$3  \\
    $\sigma_{[N II]\lambda6583}$ &[\kms ] & ~ & 103$\pm$3         & 91$\pm$36         & 88$\pm$4          & 73$\pm$2          & 89$\pm$5  \\   
    $\sigma_{[O III]\lambda5007}$ &[\kms ]& ~ & 95$\pm$1          & 98$\pm$2          & 103$\pm$3         & 86$\pm$2          & 92$\pm$3  \\    
    ~                            & ~     & ~ &    ~              & ~                 & ~                 & ~                 & ~         \\    
    vel$_{stars}$        &[\kms          & ~ & -6 $\pm$22        & 45 $\pm$12 (**)   & -25 $\pm$21 (**)  &-14 $\pm$11        & 85 $\pm$21    \\
    $\sigma_{stars}$     &[\kms ]        & ~ & 81$\pm$20        & 53$\pm$11 (**)     & 89$\pm$9 (**)      & 52$\pm$11         & 94$\pm$8  \\
    ~                    &  & ~              &    ~              & ~                 & ~                 & ~                 & ~         \\
    EBV $^{(i)}$        & [mag] & ~         & 0.27              & 0.22              & 0.45              & 0.58              & 0.26       \\
    ~                    &  & (-)            & 0.17              & 0.04              & 0.41              & 0.36              & 0.05      \\
    ~                    &  & (c)            & 0.32              & 0.30              & 0.49              & 0.65              & 0.23      \\
    ~                    &  & (+)            & 0.37              & 0.34              & 0.45             & 0.68              & 0.44       \\
    ~                    &  & ~              &    ~               & ~                & ~                 & ~                 & ~         \\    
    N$_{e_[S II]}$     &  [cm$^{-3}$]      & ~ &$70^{+0.1}_{-0.0}$&$81^{+1.9}_{-1.5}$&$298^{+3.7}_{-0.5}$&$4^{+5.2}_{-}$    &$102^{+4.8}_{-1.8}$ \\
    ~                    &  & (-)            &$40^{+0.2}_{-0.0}$ &$12^{+5.0}_{-}$    &$249^{+6.7}_{-2.7}$& -                 &$104^{+10.9}_{-0.1}$\\
    ~                    &  &(c)             &$85^{+2.0}_{-1.7}$ &$121^{+5.5}_{-2.0}$&$388^{+12.7}_{-7.9}$&$35^{+18.9}_{-20.9}$&$104^{+4.9}_{-1.0}$\\
    ~                    &  & (+)            &$80^{+2.5}_{-1.5}$ &$118^{+13.9}_{-8.2}$&$285^{+46.6}_{-38.7}$& -             &$81^{+4.9}_{-2.1}$ \\
    ~                    &  & ~              &    ~              & ~                & ~                 & ~                 & ~             \\    
    T$_{e_[S III]}$     &  [$^\circ$K] & ~   &$9495^{+65}_{-57}$   &$9729^{+90}_{-53}$    &$9118^{+72}_{-91}$    &$15437^{+86}_{-101}$   &$9767^{+39}_{-81}$ \\
        ~                &  & (-)            &$8538^{+168}_{-250}$ &$9099^{+259}_{-134}$  &$8444^{+160}_{-121}$  &$11313^{+333}_{-157}$  &$9315^{+575}_{-307}$ \\
    ~                    &  & (c)            &$10100^{+90}_{-193}$ &$10122^{+128}_{-163}$ &$9608^{+144}_{-160}$  &$16967^{+149}_{-387}$  &$9712^{+230}_{-159}$ \\
    ~                    &  & (+)            &$12590^{+151}_{-180}$ &$12498^{+161}_{-203}$ &$15321^{+193}_{-225}$ &$19152^{+596}_{-686}$  &$12602^{+281}_{-379}$  \\
    T$_{e_[O I]}$       & [$^\circ$K]      & ~  &$9149^{+5663}_{-4433}$ &$12230^{+3699}_{-2403}$ &$9702^{+691}_{-578}$ &$15620^{+3615}_{-3085}$ &$8037^{+2899}_{-2476}$  \\
    T$_{e_[N II]}$      & [$^\circ$K]      & ~  &$12544^{+3322}_{-1289}$ &$11260^{+2153}_{-1384}$    &$12408^{+1830}_{-1634}$   &$52765^{+}_{-}$   &$10776^{+2131}_{-1289}$  \\
    ~                    &  & ~              &    ~               & ~                & ~                 & ~                 & ~             \\    
    12+log(O/H)          &  & ~              & $8.50^{+0.01}_{-0.01}$ & $8.42^{+0.02}_{-0.01}$  & $8.64^{+0.02}_{-0.02}$  & $7.64^{+0.01}_{-0.01}$  & $8.51^{+0.01}_{-0.02}$ \\
    ~                    &  & (-)            & $8.73^{+0.05}_{-0.07}$ & $8.57^{+0.06}_{-0.03}$  & $8.77^{+0.04}_{-0.03}$  & $8.09^{+0.05}_{-0.03}$  & $8.67^{+0.14}_{-0.07}$\\
    ~                    &  &(c)             & $8.36^{+0.02}_{-0.04}$ & $ 8.34^{+0.02}_{-0.03}$ & $ 8.55^{+0.03}_{-0.03}$ & $ 7.46^{+0.01}_{-0.03}$ & $ 8.52^{+0.05}_{-0.03}$  \\
    ~                    &  & (+)            & $7.98^{+0.02}_{-0.02}$ & $7.98^{+0.03}_{-0.04}$  & $7.75^{+0.04}_{-0.04}$  & $7.46^{+0.05}_{-0.60}$  & $8.02^{+0.05}_{-0.06}$ \\
    ~                    &  & ~              &     ~              & ~                & ~                 & ~                 & ~         \\       
    12+log(N/H)          &  & ~              & $7.33^{+0.01}_{-0.01}$ & $7.28^{+0.02}_{-0.01}$ & $7.64^{+0.02}_{-0.02}$ & $7.01^{+0.02}_{-0.01}$ & $7.18^{+0.01}_{-0.02}$ \\
    ~                    &  & (-)            & $7.41^{+0.04}_{-0.06}$ & $7.32^{+0.06}_{-0.03}$ & $7.69^{+0.04}_{-0.03}$ & $7.35^{+0.05}_{-0.03}$ & $7.34^{+0.12}_{-0.07}$ \\
    ~                    &  &(c)             & $7.31^{+0.02}_{-0.04}$ & $7.21^{+0.02}_{-0.03}$ & $7.61^{+0.03}_{-0.03}$ & $7.03^{+0.01}_{-0.03}$ & $7.20^{+0.05}_{-0.03}$ \\
    ~                    &  & (+)            & $7.07^{+0.02}_{-0.02}$ & $7.10^{+0.02}_{-0.03}$ & $7.16^{+0.02}_{-0.05}$ & $6.87^{+0.04}_{-0.05}$ & $6.86^{+0.04}_{-0.05}$ \\
    ~                    &  & ~              &      ~             & ~                & ~                 & ~                 & ~             \\ 
    12+log(S/H)          &  & ~              & $6.53^{+0.01}_{-0.01}$ & $6.55^{+0.02}_{-0.01}$ & $6.74^{+0.01}_{-0.02}$ & $6.00^{+0.01}_{-0.01}$ & $6.65^{+0.01}_{-0.01}$ \\
    ~                    &  & (-)            & $6.71^{+0.04}_{-0.06}$ & $6.76^{+0.05}_{-0.03}$ & $6.88^{+0.04}_{-0.03}$ & $6.40^{+0.04}_{-0.02}$ & $6.81^{+0.11}_{-0.06}$ \\
    ~                    &  &(c)             & $6.52^{+0.01}_{-0.03}$ & $6.53^{+0.02}_{-0.03}$ & $6.70^{+0.03}_{-0.03}$ & $5.75^{+0.01}_{-0.04}$ & $6.62^{+0.04}_{-0.03}$ \\
    ~                    &  & (+)            & $6.21^{+0.02}_{-0.02}$ & $6.28^{+0.02}_{-0.02}$ & $6.04^{+0.01}_{-0.02}$ & $6.30^{+0.03}_{-0.04}$ & $6.33^{+0.03}_{-0.04}$ \\
    ~                    &  & ~              &      ~             & ~                & ~                 & ~                 & ~             \\    
    log(N/O)             &  & ~              & $-1.17^{+0.01}_{-0.01}$ & $-1.14^{+0.03}_{-0.01}$ & $-1.00^{+0.03}_{-0.03}$ & $-0.63^{+0.02}_{-0.01}$ & $-1.33^{+0.01}_{-0.03}$  \\
    ~                    &  & (-)            & $-1.32^{+0.06}_{-0.09}$ & $-1.25^{+0.08}_{-0.04}$ & $-1.08^{+0.06}_{-0.04}$ & $-0.74^{+0.07}_{-0.04}$ & $-1.33^{+0.14}_{-0.10}$  \\
    ~                    &  &(c)             & $-1.05^{+0.03}_{-0.06}$ & $-1.13^{+0.03}_{-0.04}$ & $-0.94^{+0.04}_{-0.04}$ & $-0.43^{+0.01}_{-0.04}$ & $-1.32^{+0.07}_{-0.04}$  \\
    ~                    &  & (+)            & $-0.91^{+0.03}_{-0.03}$ & $-0.88^{+0.04}_{-0.05}$ & $-0.59^{+0.04}_{-0.06}$ & $-0.59^{+0.06}_{-0.08}$ & $-1.16^{+0.06}_{-0.08}$  \\
   \multicolumn{3}{l}{S-calib. from \citet{Pilyugin2016}}         &      ~           & ~                 & ~                 & ~               \\
    12+log(O/H)          &  & ~              & 8.39 $\pm$0.02     & 8.33 $\pm$0.02   & 8.48 $\pm$0.02    & 8.43 $\pm$0.02    & 8.26 $\pm$0.02    \\
    ~                    &  & (-)            & 8.42 $\pm$0.02     & 8.35 $\pm$0.02   & 8.49 $\pm$0.02    & 8.47 $\pm$0.02    & 8.40 $\pm$0.02    \\
    ~                    &  &(c)             & 8.37 $\pm$0.02     & 8.31 $\pm$0.02   & 8.47 $\pm$0.02    & 8.40 $\pm$0.02    & 8.27 $\pm$0.02    \\
    ~                    &  & (+)            & 8.34 $\pm$0.02     & 8.34 $\pm$0.02   & 8.43 $\pm$0.02    & 8.46 $\pm$0.01    & 8.21 $\pm$0.02 \\
    \hline
    \multicolumn{6}{l}{$^{(i)}$\footnotesize{E(B-V) errors are $<$ 0.0037 mag.}}
    \end{tabular}
    \caption{Average physical properties of the the entire galaxy (1\arcmin\ aperture), the knots A, B, C and the star forming region of the dusty arm (1\arcsec\ aperture). In the second column, the symbols (-), (c) and (+) represent the blueshifted, central and redshifted gas. In (*) we show in addition to the average velocity dispersion, the luminosity-weighted average velocity dispersion representing the typical random motions of the ionized gas withing the galaxy. 
    Since we were unable to fit the stellar absorption with \texttt{ppxf} in knot B and part of knot A, the velocity and velocity dispersion we show (marked as (**)) were extracted from the average values of their surrounding. In the empty places in the density and temperature we were unable to estimate the values with the diagnostic used.
    }
    \label{tab:param_knots}
\end{table*}

Fig. \ref{fig:velocityHaNII} shows the velocity difference ($v_{H\alpha}-v_{[N II]}$) of the \Ha\ and [N II] emitting gas. Positive (negative) values indicate that the [N II] gas is blueshifted (redshifted) compared to the \Ha\ gas. To avoid possible contamination of the \Ha\ line to the [N II]-kinematics, we masked the broadest region in grey colour
\footnote{For the [N II] kinematics we verify that the line centre and FWHM of the \NIIr\ line are not systematically blueshifted and broadened by the close \Ha\ line, especially in the regions with broad line profiles. To achieve that, for each pixel, we constructed a model by sampling the \Ha\ and \NIIr\ lines from a combination of two Gaussian distributions that are characterised by the line centre, FWHM and \Ha\ / \NIIr\ intensity ratio parameters taken from the observations. We then performed a single fit to the \NIIr\ line in the same velocity range (-425 to 425 \kms ) as the range used in the fitting procedures to derive the kinematics. We found a maximal shift of the line centre to the blue of 0.25 \kms\ and 6 \kms\ in FWHM ($\sigma =$2.5 \kms ). Nevertheless, in the region with clearly double kinematic component we might underestimate the contamination of the \Ha\ line, therefore this area was masked in grey to avoid unreliable results there.}. 
The velocity discrepancy is considerable in some regions and range from -40 to 70 \kms, however in general terms, the [N II] gas is consistently more blueshifted. This tendency was already shown in Section \ref{section:systemicVel}, where we estimate the [N II] and \Ha\ velocities from the entire galaxy. Ignoring the masked area, the spatial distribution of the [N II] gas that is considerably more blueshifted (blue colour in Fig. \ref{fig:velocityHaNII}) interestingly draws a semi-arc structure towards the east, that encompass the starburst region. We present an interpretation of this structure in the discussion section. 

In Section \ref{sec.abundances} we estimate element abundances and we trace some nitrogen-enriched regions in magenta contours (12+log(N/H) $\geq$ 7.38) that show distinct kinematics. 
In these regions and in the red-coloured area in the south of knots A, the nitrogen-enriched gas is redshifted locally compared to the gas that recombine, suggesting at nitrogen-rich redshifted outflows.
Similar nitrogen-rich features with differential [NII]$-$ \Ha\ velocities were found in ESO 338-IG04 and were linked to Wolf-Rayet clusters within or close by these features \citep{Bik2018}. Thus, in these regions around knots A and B we are likely tracing the Wolf-Rayet nitrogen-rich wind material moving away from us.

Analysing the \OIII\ velocity field in Haro 11 we found that it agrees in most locations very well with the \Ha\ velocities.  Even so, the velocity difference ranges from -50 to 45 \kms\  in some localised regions. 
The highly ionized [O III]-gas is especially blueshifted (${\Delta}v \geq$ 20 \kms , white contours) in the southern filamentary structure and together with the [N II]-enriched gas, in the outflow close to knot B. 
In the latter, we are tracing perhaps the metal-enriched ejecta from supernovae and Wolf-Rayet stars, streaming out at higher velocities than the gas that recombines.

In the knots, the metal-enriched gas is slightly blueshifted in knot C, likely driven by supernova explosions (this is discussed in details in section \ref{sec.abundances}). In knots A and B however, the metal-enriched gas is redshifted compared to the \Ha\ gas and suggest that metal-enriched outflows from stellar winds or supernova move preferentially to the opposite direction there, perhaps due to a higher column density of gas in the side near us.

\begin{figure*}
    \centering
    \includegraphics[width=18cm]{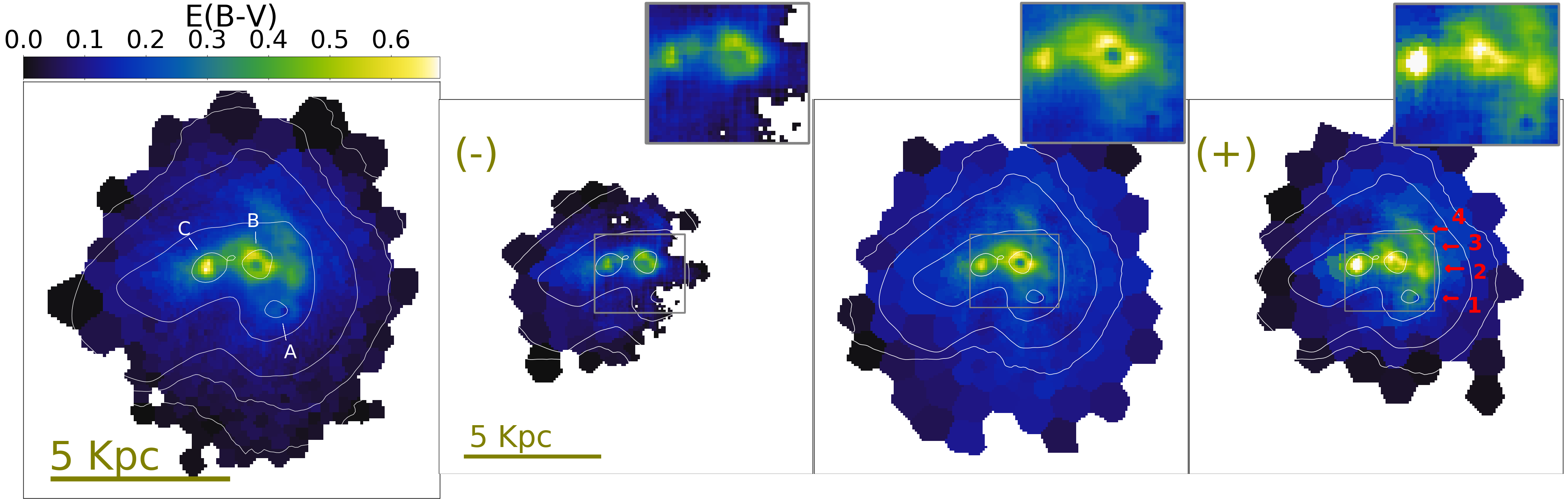} 
    \caption{E(B-V) extinction map from Haro 11. The left panel shows the nebular extinction map derived from the integrated line fluxes.  In the second, third and four column we show the nebular extinction from the gas at blueshifted (-), central and redshifted (+) velocities respectively. In the extinction of the redshifted gas (right panel) we show in red arrows the position of the four dusty lobes that form the dusty arm. The smallest panels in the top of each velocity bins display a zoom-in window from the extinction in the galaxy centre (grey box), focusing on the region around knot B. Here and in all remaining maps, the white overlaid contours shows the stellar continuum and offers a spatial orientation in the maps.
    }
    \label{fig:EBV}
\end{figure*}

\section{Properties of the warm ionized gas}

\begin{figure}
    \centering
    \includegraphics[width=1\columnwidth]{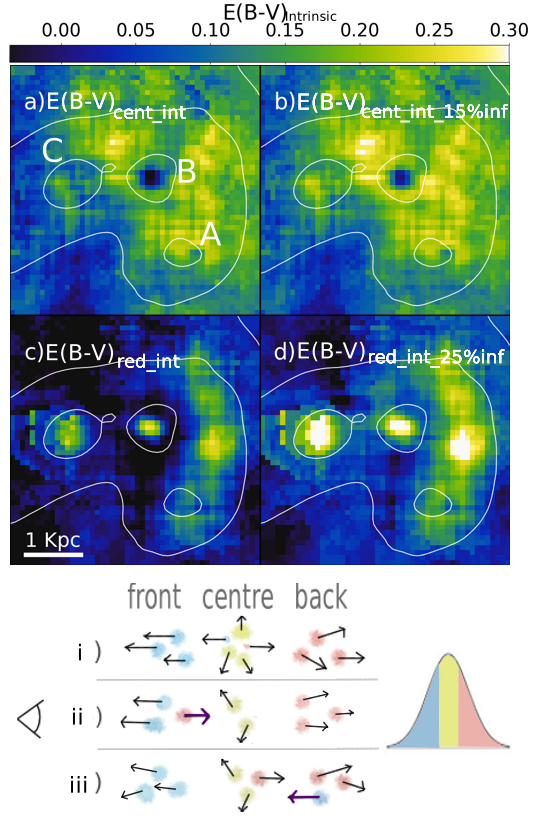} 
    \caption{Intrinsic E(B-V) values of the gas at central and redshifted velocities from the central region of the galaxy.  
    In panel \textit{a)} we show the intrinsic extinction of the gas at central velocities after subtracting the intrinsic extinction of the blueshifted gas that is in front of it. In the same manner, in panel \textit{c)} we show the redshifted intrinsic extinction obtained after subtracting the intrinsic extinction of the gas at blueshifted and systemic velocities, which is ahead of the redshifted gas.
    The black colour in both panels shows the region where we obtain negative intrinsic extinctions. Negative values are expected in case that gas inflow, in front the galaxy or from the back side, contribute to the intrinsic extinction of the gas in that particular velocity channel. 
    In panel \textit{b)} we eliminate the negative values from the very centre of knot B on condition that at least 15\% of the extinction from the blueshifted gas is due to inflows located in the back side of the galaxy with respect to the line-of-sight. This scenario is illustrated in the cartoon below the panels in \textit{scenario iii)}. 
    In panel \textit{d)} we eliminate the negative redshifted intrinsic extinction from the areas around knot B after assuming that at least 25\% of the extinction of the redshifted gas come from inflows that are located in the front part of the galaxy. This scenario is pictured in \textit{scenario ii)} in the cartoon.
    The scenario \textit{i)} in the cartoon represents the general trend of the gas in the galaxy. Most of the  approaching (receding) gas is located in front (back) of the galaxy with respect to the line-of-sight. 
    In the cartoon, blue and red colours shows the approaching and receding gas while in yellow we show the gas at central velocities. The emission of each gas component is coloured in the Gaussian curve. 
    }
    \label{fig:dust_inflows}
\end{figure}

In this section we derive the extinction, electron density and electron temperature of the ionized gas from integrated line maps and from the sequence of maps in three velocity channels. 
In Fig. \ref{fig:EBV} to \ref{fig:temperaturee} we show results from the integrated maps in the large panel on the left and results from the gas at blueshifted "(-)", central, and redshifted "(+)" velocities in the remaining smaller panels.

\subsection{Extinction and dust in Haro 11}\label{"subsec:dust"}

The nebular extinction maps are derived from the stellar absorption corrected \Ha\ and \Hb\ lines, assuming case B recombination (T$=$10$^4$ K and Ne= 100 cm$^{-3}$) and the Cardelli extinction curve \footnote{Using the SMC extinction law does not change our results} \citep{Cardelli1989}. 
The Galactic extinction, although very low ($\sim$0.01 mag), was subtracted from the final maps. 
Fig. \ref{fig:EBV} shows the derived E(B-V) maps towards Haro 11. 
The nebular extinction is particularly elevated in knot C, in the region of knot B, and in the dusty arm at redshifted velocities. While in knot C it is mostly confined in the compact core ($<$1\arcsec), the largest values in knot B are tracing an intriguing ring-like structure. 
This dusty ring, which is visible at central velocities\footnote{Given that the \Ha\ map was convolved to match spatial resolution of the \Hb\ map, we tested if this procedure has artificially created the dusty ring. By extracting a E(B-V) from the original (non-convolved) \Ha\ and \Hb\ maps we found the same ring structure, suggesting that this dusty ring is real.}, is about 430 pc in radius (1 \arcsec) and the E(B-V) difference from the ring to its low-dust or dust-free interior is about 0.2 mag. This structure strongly hints at a shell of dust formed by the young cluster population in knot B (average age $<$ 5 Myr). Many of these clusters are very massive (M$_{cl}>10^5$ M$_\odot$), containing up to thousands of massive stars \citep{Adamo2010}. Hence, it is reasonable that radiative pressure and stellar winds were the main mechanisms that have cleared the core of knot B from dust, and formed the dusty shell.
A similar, but more diffuse dusty shell (r$\sim$210 pc) is seen in knot A (or perhaps the first lobe of the dusty shell) at central and redshifted velocities. 
IR bubbles formed from PAHs and dust, seems to be common in star forming regions where massive stars form. \citet{Churchwell2009} catalogued hundreds of closed and open IR bubbles in the Milky Way plane from SPITZER observations. Some of these bubbles were found to have a cavity at 24 $\mu$m (a dust tracer), indicating that the dust inside have been evacuated \citep[e.g.,][]{Watson2008,Zhang2013}. A somewhat similar scenario seems to occur in the core of knot B and probably knot A. 
In the extragalactic field, it has been difficult to detect IR bubbles because of the low spatial resolution of the SPITZER instrument, but this is expected to change with the upcoming JWST telescope.

In the dusty arm, on the other side, the clumps of dusty gas seems to compact in four oblate sub-regions with ongoing star formation (red arrows in fig. \ref{fig:EBV}). From south to north: the first lobe coincides with knot A towards the west and was disentangled only in the kinematics (${\Delta}v=$40 \kms ), the second lobe is the most dusty, the third is the brightest in the optical spectra of all and hosts several young star clusters \citep{Adamo2010} and the northern one is elongated and more diffuse. In table \ref{tab:param_knots} we show the physical properties of the brightest (third) star forming lobe northwest of knot B that is marked in Fig. \ref{fig:HaFlux}. 
Beyond the starburst region, the extinction is rather low (E(B-V)$\leq$0.15 mag) and decreases monotonically with radius.

Comparing the extinction of the gas at different velocities, the E(B-V) values increase from blueshifted to redshifted velocities. This differential extinction can be used to constrain the location of the gas at the different velocities, because the gas in the near side of the halo will have lower extinction than the gas at the far side, as light emitted at the far side is travelling through a larger column of gas and dust. 
Our lowest extinction values in the blueshifted gas imply that a considerable fraction of the approaching gas is on the near side of the galaxy, and the redshifted gas on the far side. 
The extinction values presented in table \ref{tab:param_knots} show that this is the case for the galaxy as a whole, the knots and the dusty arm with one exception: In knot B the extinction from the redshifted gas is lower than that of the gas at central velocities. This would then suggest that a considerable fraction of the redshifted gas in knot B, is located on the near-side of the galaxy. This effect is expected in cases where inflows contribute to the extinction. We investigate this in more detail in next subsection.

Comparing the extinction values in the knots from the extinctions derived in the same manner by other authors, we found in general better consensus for the extinctions in knots A and B (E(B-V)$\sim$ 0.22 and $\sim$0.45 mag resp.) than in knot C (E(B-V)$\sim$ 0.58) (see table \ref{tab:param_knots}).  We note that the extinction and other physical parameters derived from knot A might include emission from the dusty lobe overlying knot A, especially at redshifted velocities. 
For knot C, \citet{hayes2007} and \citet{James2013} found an extinction of 0.42 and 0.48 mag while \citet{guseva2012} reported a very low value of 0.18 mag. The reason for that is that \citet{guseva2012} subtracted a fraction of the knot C \Ha\ flux that they attributed to luminous blue variable stars. Without this correction, they would have derived an extinction of about 0.62 mag.

\subsubsection{Gas inflows}

We investigate possible inflows from the intrinsic extinction of the gas in the three velocity channels. 
The cartoon in Fig. \ref{fig:dust_inflows} present three possible scenarios. Blue, yellow and red colours show the position of the gas emitting at blueshifted, systemic and redshifted velocities, as seen in the Gaussian curve. The arrows represent the velocity vector. 
In scenario \textit{i)} we illustrate the general configuration of the gas in Haro 11 with respect to the line-of-sight: the approaching gas is located in the front and the receding gas in the back side. 
In scenario \textit{ii)} we picture an inflow located on the front side (violet arrow), which has redshifted velocities. 
This inflow will have a lower extinction than the gas behind it and will lower the average extinction of the redshifted gas. If the dust content in the inflow is significant relative to the dust at central velocities, the redshifted gas will then have a lower extinction than that of the gas at central velocities.
In scenario \textit{iii)} we illustrate an inflow located in the back side of the galaxy with respect to the line-of-sight (purple arrow), which has blueshifted velocities. This scenario will result in a higher extinction of the blueshifted gas than that of the gas at central velocities. 

\begin{figure*}
    \centering
    \includegraphics[width=2\columnwidth]{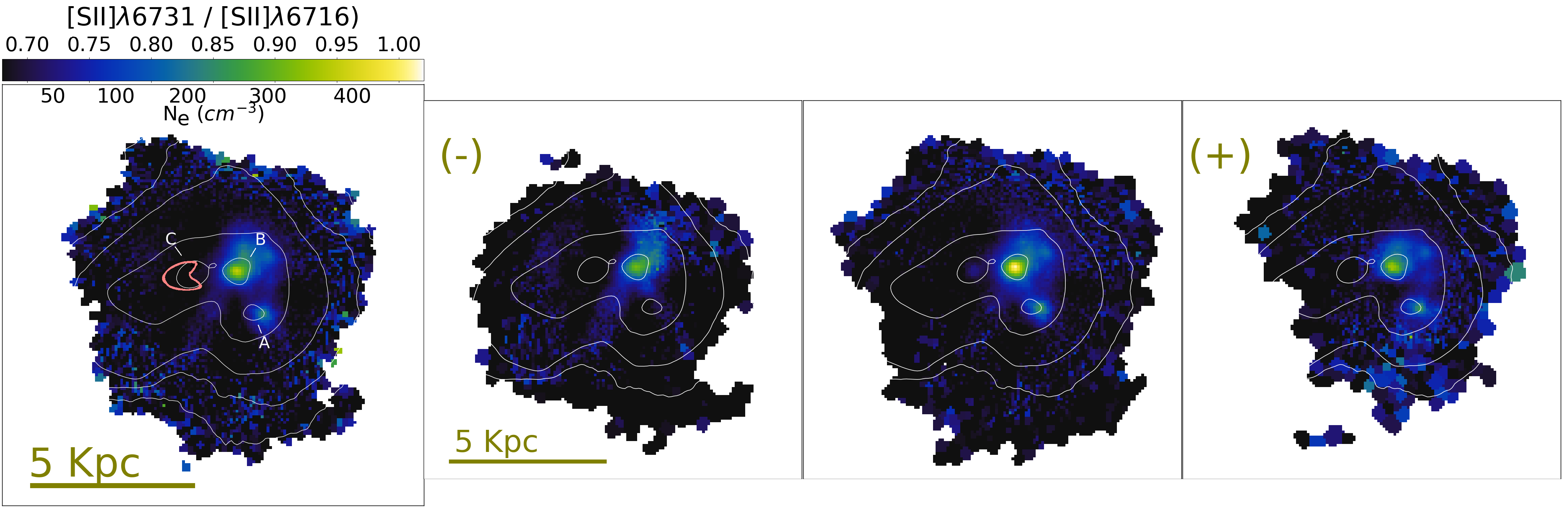}
    \caption{\SIId\ line ratio maps used to compute the electron density of the gas. The colorbar shows both the line ratio above and the corresponding density below the bar.
    We show in the large panel the density from the integrated lines and in the smaller panels from left to right the density of the blueshifted, central and redshifted gas. The pink contour in the left panel shows the region where the [S III] line ratios are lower than the minimum value of $\sim$0.69 required to estimate a density N$_{e}>$ 1 cm$^{-3}$ with this tracer. 
    The density is higher in knot B, A and the dusty arm while it is extremely low in knot C and around it. }
    \label{fig:density}
\end{figure*}

The intrinsic extinction of the gas at central (E(B-V)$_{cent\_int}$) and redshifted (E(B-V)$_{red\_int}$) velocities are obtained by subtracting the intrinsic extinction of the gas in front of it. For the blueshifted gas, the intrinsic extinction is the same as the measured one, as it is the closest to us.

\begin{equation}\label{eq:ebminv_sys}
\begin{split}
E(B-V)_{cent\_int} &= E(B-V)_{cent} - E(B-V)_{blue}\\
(E(B-V)_{red\_int} &= E(B-V)_{red} - E(B-V)_{cent\_int} - E(B-V)_{blue} \\ 
& =  E(B-V)_{red} - E(B-V)_{cent} &
\end{split}
\end{equation}

Intrinsic values greater than zero will confirm the simple gas-dust geometry seen in scenario \textit{i)}, otherwise if these are negative this simplistic scenario does not hold anymore and part of the redshifted/blueshifted gas must be infalling. We can then calculate the minimum E(B-V) contribution of the inflow based on what would be required to eliminate the intrinsic negative E(B-V) values. We note that this contribution could, in theory, be larger.

Fig \ref{fig:dust_inflows} shows in panels \textit{a} and \textit{c} the intrinsic extinctions of the gas at central (E(B-V)$_{cent\_int}$)  and redshifted velocities (E(B-V)$_{red\_int}$). 
We find in general intrinsic extinctions greater than zero confirming scenario \textit{i)}. There are, however, some places with negative values suggesting inflows in the region of knot B (scenarios \textit{ii} and \textit{iii}), towards the north of knot C and nearby the broad line region with two kinematic components (scenario \textit{ii}).

We zoom-in on the region of knot B and perform a further analysis on the dusty shell. 
The dusty shell, as indicated by the dusty ring, has a diameter of  \diameter $=$  2.2\arcsec\ $\sim$ 950 pc and a thickness of dr $=$ 1\arcsec $\sim$ 430 pc. The interior of this shell has an approximate diameter of \diameter $=$ 0.8\arcsec $\sim$ 340 pc. We additionally analyse the very centre of the bubble with a size of \diameter $=$ 0.4\arcsec $\sim$ 170 pc, smaller than the seeing (0.8\arcsec ).
We compute the average measured and intrinsic extinctions from the shell area and the shell interiors, which are tabulated in table \ref{tab:EBV_int_dusty_bubble_knotB} in the appendix. 

We find that the bubble's interior is effectively dust-free at central velocities (E(B-V)$_{cent\_int} =$ 0.006 mag). The blueshifted and redshifted gas is on the other side more extincted, tracing perhaps part of the dusty shell that is in the nearest and farthest side. 
We note nonetheless that the very centre of the bubble has an average intrinsic value of E(B-V)$_{cent\_int} =$ $-$0.04 mag at central velocities, suggesting an inflow from the far side (scenario \textit{ iii)}. 
These negative values would vanish (avg. value become zero) if we assume that 15\% of the blueshifted extinction\footnote{E(B-V)$_{cent\_int} =$ E(B-V)$_{cent} -$ 0.85 (E(B-V)$_{blue}$)} in the regionof knot B is due to an inflow from the back side. 

In the dusty ring, on the other hand, we measure in average a negative intrinsic extinction of the redshifted gas (E(B-V)$_{red\_int} =$ $-$0.028 mag), suggesting inflows from the near side of the galaxy (scenario \textit{ ii)}. \citet{Ostlin2021} present further evidence of inflows in the knots from the analysis of absorption features in HST/COS UV spectra. The authors found substantial Si II and Si IV absorption in the gas at v$\ge$75 \kms\ towards knot B, suggesting that there is neutral and ionized receding gas in the front side (inflows) with respect to the line-of-sight. In knot A the absorption lines are somehow shallower suggesting that there less redshifted gas is falling in; and in knot C redshifted absorption features were found mainly in the ionized gas. Thus, the redshifted intrinsic extinction in knot B is likely affected by the extinction of the infalling gas traced by these authors.
If we assume that 25\% of the intrinsic redshifted extinction\footnote{
E(B-V)$_{red\_int} =$ E(B-V)$_{red} -$ 0.75 (E(B-V)$_{cent}$)} in knot B region comes from this infalling gas (illustrated in Fig. \ref{fig:dust_inflows} \textit{d}), we obtain E(B-V)$_{red\_int} =$  0, within the errors, in the dusty ring. The 25\% fraction is only an lower limit, since the dusty ring might have E(B-V)$_{red\_int} >$ 0 mag.

Note that our estimates are in general suited for a dust screen model, where the dust is more uniformly distributed. However, this values will change if the dust has a non-uniform clumpy distribution. In that case, it will decrease the E(B-V) compared to the more homogeneous screen model \citep{Natta1984}.

\subsection{Electron density}\label{"subsec:density"}

We derive the electron density maps from the \SIIr\ and \SIIb\ line ratio using \texttt{Pyneb} \citep{Luridiana2015}. These doublets are affected by telluric absorption, 
and although this was corrected, it was not possible to recover the line fluxes with fidelity, therefore to mitigate this effect, we bin the data using Voronoi tesselations (S/N$=$20), and use only the bins with SNR$>$25 and a minimal flux of the weakest \SIIr\ line of 25$\times 10^{-20}$ erg s$^{-1}$ cm$^{-2}$.

Fig. \ref{fig:density} shows that the electron density in Haro 11 is in general very low, with typical values less than 50$\pm$8 cm$^{-3}$. The dense gas is concentrated mostly in knot B, followed by knot A and the star forming lobe in the dusty arm and the bright outflow close to knot B (see table \ref{tab:param_knots}).  
In contrast to them, in knot C and around it we measure the lowest density in the galaxy (N$_{e}<$10 cm$^{-3}$). The left panel of fig. \ref{fig:density} shows in pink contours the area where the [S II] ratios are even lower than the minimum value ($\sim$0.69) needed to compute the densities for N$_{e}>$ 1 cm$^{-3}$. This area encloses knot C at blue- and redshifted velocities. Thus, the extreme low densities seems to enclose a cavity around knot C. The somewhat higher densities on the west of this knot most likely belong to the star forming region in knot B. This is consistent with the assumption of an empty bubble around knot C that was created by the intense feedback of its massive star cluster population \citep{menacho2019}.
Comparing these values with the densities derived from the [O II]$\lambda$3726,3729 lines by \citet{James2013}, our values are lower in all three knots. These authors derived densities of about 400 cm$^{-3}$ for knot B and A and 150 cm$^{-3}$ for knot C. 
The densities estimated from these ions normally agree well in HII regions, however the telluric absorption in our [SII] lines, although corrected, could still be an issue here. Some differences could arise also from the partially different phases of the S$^{+}$ and O$^{+}$ ions.

\begin{figure*}
    \centering
    \includegraphics[width=2\columnwidth]{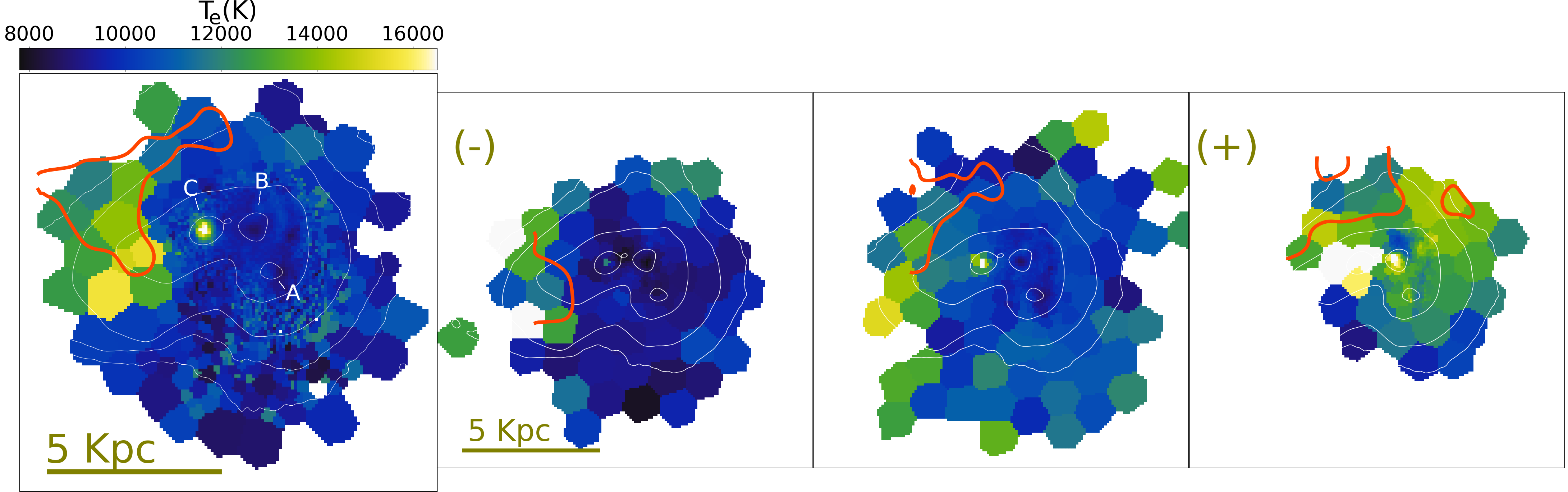}  
    \caption{Electron temperature maps derived from the \SIIIb\ and \SIIIr\ lines. 
    We show temperatures from the integrated lines in the large panel (range from 8500 to 18000 K, flux-weighted avg T$=$10000$\pm$60 K), and temperatures from the blueshifted, central and redshifted gas from left to right in the small panels. 
    Red contours show the places where \OIHa\ $>$0.1. High ratios (>1) are common indicators of fast shocks \citep[][]{Allen2008}. 
    The temperatures in the star forming knots and the halo are not homogeneous overall. Integrated temperatures of 15000 K are found in knot C and the areas that overlap with the shocked regions in the halo. Lower values were measured in the remaining star forming knots ($\sim$9500$\pm$75 K) and the non-shocked halo ($\sim$10000$\pm$373 K). The luminosity-weighted average temperature of the galaxy increase from blueshifted to redshifted velocities (T$_{blue}\sim$8500$\pm$300 K, T$_{cen}\sim$10100$\pm$250, T$_{red}\sim$12600$\pm$165).
    }
    \label{fig:temperaturee}
\end{figure*}

\begin{figure}
    \centering
    \includegraphics[width=\columnwidth]{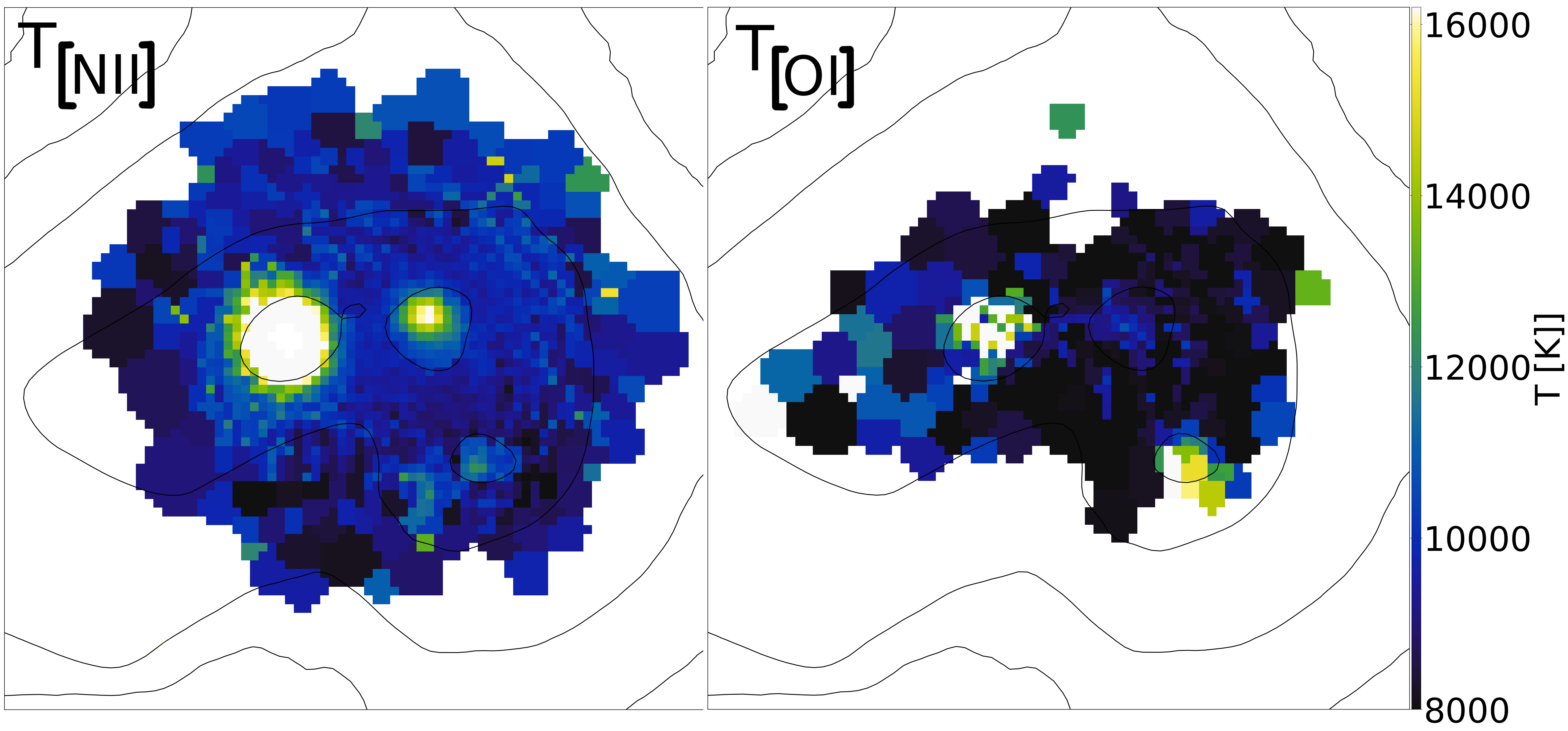}
    \caption{[N II] and [O I]-temperatures from the central starburst region. The [N II]-temperature in the knots are in general higher than the values estimated from the [S III] lines. In Knot C, the temperature measured is $\sim$52760 K, which is likely overestimated due to recombination excitations. Taking into account this contribution, the the new estimated temperature is about 16800 K. 
    The [O I]-temperatures are similar to the [S III] values, except for knot A that is highly ionized.
    }
    \label{fig:temp_OI_NII}
\end{figure}

\subsection{Electron temperature}\label{"sec.temp"}

The electron temperature maps are derived from the extinction corrected \SIIIb\ auroral line and \SIIIr\ nebular line. The blue part of these lines were integrated till -125 \kms to avoid contamination to the auroral line from the  close \OI\ line. 
Aiming to derive reliable temperatures to the farthermost part of the halo, we Voronoi binned the weak auroral line with a maximum cell size of 200 pixels ($\sim$1.5 kpc$^2$) and masked the regions with SN$<$6 (see Appendix Fig. \ref{fig:SIII6312_line}). The pattern was then applied to the redder [S III] line and the line ratio was calculated. We use the task \textit{getTemDen} from \texttt{Pyneb} that computes the temperature once the density is provided. This method finds a self-consistent result by cross-converging the temperature from the density and the ratio of the temperature-sensitive lines. 

Fig. \ref{fig:temperaturee} shows the electron temperature up to a distance of 7 kpc in radius. 
The red contours mark the regions affected by fast shocks that were outlined in \citet{menacho2019}. 
At first glance, the temperature distribution in the halo appears bi-modal. 
In the large north-east part, the temperatures are considerably higher, but in most part of it the integrated temperatures oscillate between 8800 and 11500 K with an average value close to 10000 K. 
The errors oscillate from 200 K in the central regions to a maximum value of 500 K in the outskirts.
In the halo there are clearly temperature variations in contiguous cells that get diluted, but do not disappear if we, for example, increase considerably the signal-to-noise threshold of the Voronoi binning, which increases the sizes of those small cells in the halo (Using a S/N$=$50 instead of the S/N$=$6 used in the maps of Fig \ref{fig:temperaturee}, the temperature in the halo excluding the hot north-east part oscillate from 9100 to 10800 K). However, these variations can be washed out if the cell sizes increases considerably. This imply, that possible temperature (and metallicity) variations are more diluted in the outskirt of the galaxy where we have larger cells. 

Most of the regions with lower temperatures belong to the regions where we see double kinematic components, the filamentary structure (even at blueshifted velocities) and the western ionized cone. 
Given that the filamentary structure is highly ionized, the low temperature values are perhaps surprising. 
An explanation could be, that, these low temperatures might arise out from gas condensations in the filaments, but not from the rarefied and potentially hot gas between the filaments. 
We note however that the temperature of this region would be better traced from the [OIII] line that arises from these highly ionized environments.

In the area with the highest temperature in the halo (NE region in the integrated map), the estimated values reach up to 15000$\pm$500 K. This area (about 7$\times$3 kpc$^2$) clearly coincides with the regions where fast shocks traced from the \OIHa\ ratio are detected \citep{menacho2019}. Here, the gas is characterised by a low density (see Fig. \ref{fig:density}), a low ionization and a bright diffuse \La\ radiation that originates from scattering of \La\ photons in a neutral medium \citep{hayes2007}. \citet{Bergvall2000} have previously pointed out the impact of shocks in the galaxy.  \citet{menacho2019} in turn suggested that supernova-induced shock waves from knot C are likely propagating through the ISM of this structure. \citet{Peimbert1991} have shown that shocks tend to enhance the strength of the auroral lines considerably. Thus, shocks could be potentially affecting our estimated temperatures here, which deviate up to 5000 K from the average halo temperature. 
To our knowledge, such large-scale areas in the halo of other galaxies where shocks might have considerably enhanced the derived temperatures have not been found previously. The likely reason is the weakness of the auroral lines that require very deep observations with long integration time. Evidence of the impact of shocks in the derived temperatures from collisional excited lines come mostly from planetary nebulae where stellar winds seem to shock the halo of these objects \citep{Middlemass1989, Esteban2002}.

Comparing the temperatures in the different velocity panels, this clearly increases from the blueshifted to the redshifted gas (${\Delta}T\sim$4000$\pm$300 K).
In all these velocity bins, the enhanced temperature cells lie in and nearby the shocked area.
At central velocities we note that the highly ionized gas streaming out from knot A towards the southern filamentary halo hat a uniform temperature of about 11000$\pm$280 K, which is slightly hotter (${\Delta}T\sim$1000$\pm$500 K) than its surrounding gas. Perhaps we are tracing a larger fraction of the highly ionized hot gas that is not traced at blueshifted velocities.

The gas condition in the knots are significantly different. 
Knot C has exceptionally low densities and high temperatures while knot B is denser and cooler. Knot A and the dusty arm on the other side have typical temperatures of HII-regions (see Table \ref{tab:param_knots}). 
Similar to the gas in the halo, in all star forming condensations the electron temperature increases from the blueshifted to the redshifted gas. 

The temperatures in the knots were previously derived by \citet{guseva2012}, \citet{James2013} and \citet{Cairos2015} using [O III]-temperature sensitive lines from long slit or IFU data from different instruments. The [O III]-lines arise from a gas that is more ionized than the gas where the [S III]-lines are emitted. Despite that the auroral [O III]$\lambda$4363 line is not covered by MUSE at the redshift of Haro 11, we have derived the [O III]-temperatures using the equations presented in \citet{Garnett92}. These are for knots A, B and C: 9673, 8990 and 16365 K, with typical errors $\pm$100 K. 

We derived the [OIII]-temperatures for knots A and B also from the T$_{[SIII]}$ - T$_{[OIII]}$  relation from \citet{Berg2020}, who recently found from observations of HII regions that the relation from \citet{Garnett92} can deviate significantly in the highly ionized zones, especially at the low and high temperature end. We find that these temperatures are only 100 and 300 K higher than the values obtained using the relation of \citet{Garnett92}.  

Our temperatures are similar or 2000 K lower for knots A and B than the values found by \citet{guseva2012}, \citet{James2013} and \citet{Cairos2015}. For knot C however, there is no systematical trend. While \citet{Cairos2015} derived up to 2000 K higher temperatures, this is about 5000 K lower in the work of \citet{guseva2012}, who derived similar values for the [S III] and [O III]-temperatures. We point out however, that these authors used much greater apertures for the knots than our 1\arcsec\ apertures. Using larger apertures in our work would result in higher temperatures for knot B, but lower values for knot C. Our estimations are derived for the central star forming region, while the previous authors enclose also gas from the knots surroundings.

\subsection{[N II] and [O I] temperatures}

We derived additionally the [N II]- and [O I]-temperatures from the [N II]$\lambda$5755/\NIId\ 
and [O I]$\lambda$5577/\OI\ temperature-sensitive lines respectively (see Fig. \ref{fig:temp_OI_NII} and table \ref{tab:param_knots}). The weakness of these auroral lines have constrained the measurements only to the central starburst. 

The [N II]-temperatures have larger uncertainties, but are in general higher than the [S III]-temperatures in the knots and similar in the knots surroundings. 
A particular case is knot C, where the derived temperature of 52700 K is very much higher than the $\sim$15000 K derived from the [S III] and [O I] lines. As a consistency check, we also derive the [N II]-temperatures from the line fluxes published by \citet{guseva2012}, that were obtained from the VLT/X-shooter instrument. We obtain similar temperatures for knot B, but a temperature of 40000 K for knot C (or 46000 K when using our extinction value).
Thus, both data-sets give extreme temperatures for knot C. We note however that we were unable to derive uncertainties for this knot. In this range of temperatures, the [N II]-line ratios are just in the top limit of the ratios that provide temperature estimations.

High [N II]-temperatures -- excesses up to 2000 K -- are frequently derived in planetary nebulae whose nitrogen atoms are mostly in a double ionized state. This discrepance is attributed to the contribution of recombination excitations to the collisionally excited auroral line\footnote{In some elements, the low-lying metastable levels are not only excited by electron collisions, but also from recombination processes. \citet{Rubin1986} found that this effect take place due to the large effective dielectronic recombination coefficient to this levels and due to direct radiative recombinations. This effect will be high if these low-lying transitions occur in the highly ionized gas; i.e. along the line-of-sight that penetrate deeper in the [O III]-dominant interior of an HII-region.}, which consequently overestimates the real [N II]-temperatures \citep{Rubin1986}. \citet{Liu2000} provided a simple equation to calculate this contribution from the double ionized nitrogen abundance (N$^{2+}$/H$^{+}$). We were unable to derive this due to the absence of a double ionized nitrogen line in our data, but results on the oxygen abundances in knot C (Section \ref{sec.abundances}) suggest an overpopulation of double ionized species. There is, however, an indirect way to get an approximate value. For pure recombination excitation, the \NIId\ / [N II]$\lambda$5755 ratio is 6.3 for T$_e =$ 15000 K \citep{Liu2000}. The observed ratio in knot C is 12.92, which is a factor of 2 larger; thus the effects of recombination excitation on the auroral line seems to be about 49 per cent. If we subtract this contamination from the auroral line, we get a [N II]-temperature of 16815 K, similar to the [S III]- and [O I]-temperatures within the errors. 

The contribution of recombination excitation is even stronger in the \OIId\ auroral lines and therefore can also be estimated from the double ionized oxygen abundance (O$^{2+}$/H$^{+}$) \citep{Liu2000}. Contrary to our previous result, we find a contamination of only $<$ 5 per cent in knot C. Thus, we find in knot C a considerably excess in the strength of the [N II] auroral line only. We discuss in section \ref{subsec:temp and abund variations} this discrepancy from the analysis of the gas condition in knot C.

\begin{figure*}
    \centering
    \includegraphics[width=2\columnwidth]{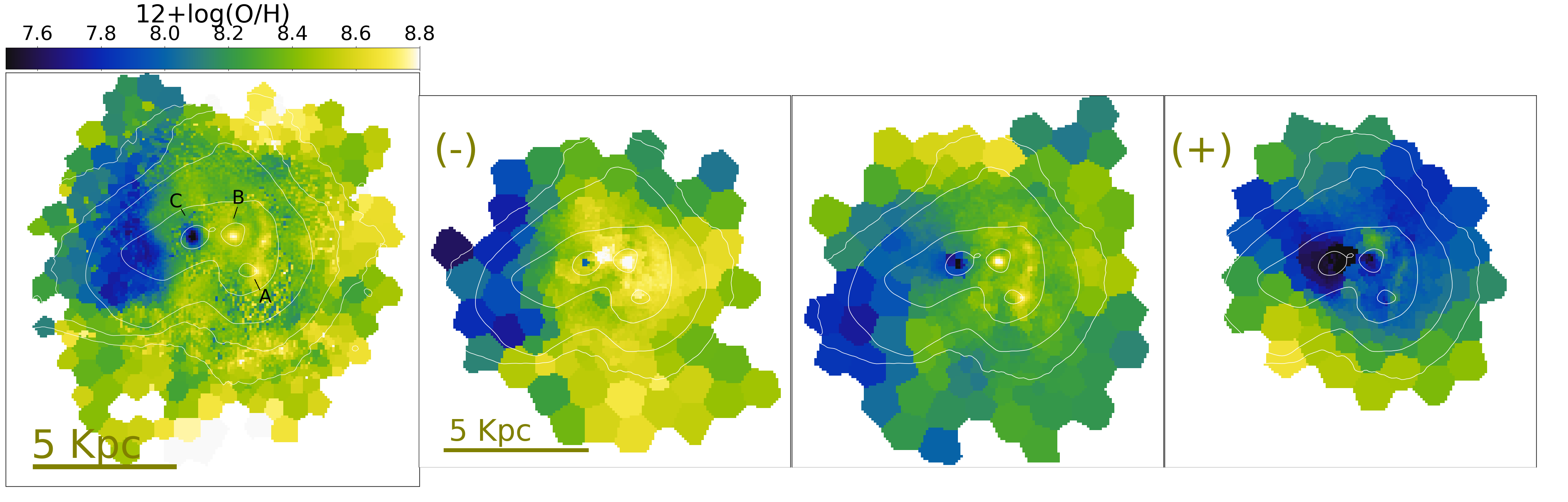}
    \caption{Oxygen metallicities derived from the direct method. The top panel shows the integrated metallicities while the three smaller panels in the second row display the metallicities of the blueshifted (-), central (c) and redshifted (+) gas. }
    \label{fig:element_abundance}
\end{figure*}

For the whole galaxy, the [N II]-temperature is about 3000$\pm$2366 K higher than the value derived from the [S III] lines. \citet{Loaiza2019} recently found [N II]-temperature discrepancies up $\pm$5000 K in four Lyman Break Analogs. Thus, higher [N II]-temperatures might indicate extreme conditions in some regions of the galaxy.

The [O I] lines on the other side trace a gas that is mostly neutral or lowly ionized. The [O I]-temperatures are similar to the [S III]-temperatures withing the errors.

\section{Metallicities} \label{sec.abundances}

Chemical abundances provide crucial information about the evolution of galaxies. They are strongly linked to the past star formation, the mass inflows and outflows and the mixing processes of metals in the ISM.  Nowadays there are three main techniques to derive chemical abundances in the optical range. The first dubbed "direct method" or "T$_{e}$-method" use direct measurements of the temperature in combination with line ratios of forbidden or recombination lines relative to the \Hb\ Balmer line. The second use photoionization models to interpret the strength of the principal lines and so to constrain the metallicities. The third is called the "strong line method" and use bright forbidden lines relative to the \Hb\ that are calibrated based on photoionization models or empirical relations derived from regions where the direct method has been used. 
In this work we analyse and compare Haro 11 metallicities from the direct method using collisionally excited lines and a strong line method.

\subsection{Metallicities derived from the direct method}

Probably the most critical part in determining the element abundances in the direct method, is the estimation of the temperatures from the gas these ions populate. 
We consider a multi-phase medium with three zones: a low ionization zone for the O$^{+}$, N$^{+}$ and S$^{+}$ ions, an intermediate for S$^{2+}$ ions, and a high ionization zone for O$^{2+}$ ions. 
We start with the T$_{[S III]}$ \footnote{Prior to the temperature estimations of the low and high ionization zones, the T$_{[S III]}$ maps from the halo were smoothed to have softer transition between voronoi-cells. To do so, we masked the central starburst (SN>10 and cell area < 4 arcsec$^2$ in the \SIIIb\ line) and convolved the T$_{[S III]}$ maps with a Gaussian kernel of FWHM$=$1.2 arcsec for the integrated line temperatures or FWHM$=$1.8 for the temperatures of the gas in velocity bins. These kernels were chosen to be smaller than the size of the halo cells, so that the temperature imprints are not washed away. } and compute the temperatures of the low and high ionization zones following \citet{Garnett92} prescriptions. 
Here we determine the metal abundances of oxygen, nitrogen and sulphur following the formulation of \citet{Izotov2006}, whose empirical expressions are based on SDSS data of metal-poor galaxies. Their ionization correction factors (ICF), which are intended to account for the abundances of the ions not covered by the observations, were determined from photoionization models. In absence of the [O II]$\lambda$3727,29 lines, we estimate the O$^{+}$ abundances from the \OIId\ doublets. Other lines used are the \OIII , [O III]$\lambda$4959, \NIId , \SIId\ and \SIIIb\ lines, with their associated densities and temperatures for the calculations of the O$^{2+}$, N$^{+}$, S$^{+}$ and S$^{2+}$ abundances respectively. The element abundances were then computed by adding the(their) ion(s) abundance(s) and, if applicable, their associated ICF.  
Here, likewise the previous ISM parameter estimations, the integrated and individual channel maps from the input parameters (here: line ratios, density and temperatures) were used for the estimations of the corresponding integrated and channel maps of the abundances.
In the appendix, we show in the Fig. \ref{fig:ratios} four line ratios from the integrated maps that were used to derive the integrated a) temperature, b) and d) O$^{+}$ and O$^{2+}$ abundances and c) N$^{+}$ abundance. These maps provide  valuable support in the interpretation of the element abundances. 

A problem that can arise in the highly ionized zones, is that, the \OIId\ auroral lines can be additionally excited by recombinations of O$^{2+}$ ions, as pointed out in the [N II]-Temperature estimations. \citet{Liu2000} provided a simple equation to estimate this contribution from the double ionized oxygen abundance (O$^{2+}$/H$^{+}$). We found this contribution to be low; it is up to 6 per cent (12+log(O/H) $\leq$ 0.02) in knot A, both ionized cones and in the filamentary structure, nevertheless, this was subtracted from the estimated oxygen abundances.

Fig \ref{fig:element_abundance} shows the oxygen metallicity maps up to a distance of about 7 kpc in radius. There are clearly inhomogeneities in the oxygen abundances across the galaxy. We measure average metallicities of 12+log(O/H) = 7.94$\pm$0.085 in the shocked area east of knot C, but in the remaining parts of the halo, the metallicities are considerably higher, with average values of 8.44$\pm$0.054 for average temperatures of 10000 K. In both cases the average standard deviation is about 0.016. In the halo, the highest metallicities (avg. 8.60$\pm$0.06) are measured in the highly ionized cones and part of the filamentary structure, though these values are perhaps overestimated since the [O III]-temperatures in these zones are better estimated from the T$_{[SIII]}$ - T$_{[OIII]}$ relation from \citet{Berg2020}. Using their relation, the T$_{[OIII]}$ increases in average 350 K and the metallicity decreases accordingly by 0.1$\pm$0.02 dex.

Comparing the abundances in the halo at different velocities, the blueshifted gas has the highest oxygen abundances (avg. 8.50$\pm$0.04) and the redshifted gas the lowest (avg. 8.15$\pm$0.06).
There are two possible explanations for the contrasting metallicities: a) the different metallicities of the galaxy progenitors; and b) a significant contribution of metal-enriched supernova ejecta in the approaching gas. This assumption is founded by observations of strong blueshifted UV absorption lines in the knots underlaying the presence of powerful outflows \citep{rivera-thorsen2017,Ostlin2021}. 
We crudely estimate the metallicities of the galaxy progenitors from the well-disentangled doubly peaked broad line region highlighted in the $\sigma$-map (region 1), where the kinematic components of both galaxies overlap. Assuming that each of these components trace the gas of one progenitor, we find the blueshifted progenitor (avg. 12+log(O/H) $=$ 8.47$\pm$0.03) to be more metal enriched by 12+log(O/H) $=$ 0.3$\pm$0.06. This difference is similar, within the errors, to the metallicity discrepancy of the blue- and redshifted gas. Thus, the contribution from metal-enriched supernova ejecta in the blueshifted gas might be from low to insignificant.

In the knots the oxygen abundances are markedly different (see Table \ref{tab:param_knots}). 
Knot B with a metallicity of 8.64$\pm$0.02 is the most metal-rich region from the galaxy. 
On the opposite side, knot C has the lowest oxygen abundances, with integrated values in average 1$\pm$0.02 dex lower than the value measured in knot B (or 0.9 dex when using T$_{[SIII]}$ - T$_{[OIII]}$ relation from \citet{Berg2020} in knot B).
In Fig. \ref{fig:ratios} of the appendix, we see indeed that (independently of the temperatures) the line ratios used to derive the abundances ([N II]/\Hb , [O II]/\Hb\ and [O III]/\Hb ) are relatively low in knot C and higher in knot B. We will discuss in section \ref{section:n_over_o} possible scenarios explaining the metal abundances in the knots. Comparing the oxygen abundances at different velocities we find that in the knots, similar to the halo, the blueshifted gas is considerably more metal-rich than the redshifted gas.
For the entire galaxy we derive a metallicity of 12+log(O/H) $=$ 8.5 from the integrated lines and the same trend of decreasing metallicities towards the redshifted velocities.

Comparing our metallicities in the knots with the values presented in the \citet{guseva2012} and \citet{James2013}, we find larger discrepancies for knots B and C. For knot A, our metallicity is only slightly higher. The reason is likely due to the different aperture sizes that each of these author have used. \citet{guseva2012} used a 5 times larger aperture (in area) than ours. The apertures used in  \citet{James2013}, are about 8, 14 and 3 times larger for knot A, B and C respectively, than our 1\arcsec\ apertures. A larger aperture can considerably affect the results in the sense that the line ratios, from which the physical conditions are derived, are computed from the total fluxes inside the aperture. A large aperture will dilute any contrasting values (e.g. temperature and metallicity) from a smaller region inside the aperture. The metallicities are doubly affected because of the strong dependence on the temperature, whose values are already averaged by the size of the aperture, and the lines ratios from which the metallicities are derived. 
In fact, we have used the same aperture as \citet{James2013} in knot C and derived an oxygen metallicity of 12+log(O/H) $=$ 7.9$\pm$0.02 which is similar, within the errors, to their value of 7.8$\pm$0.13. The high and low metallicity values we measure are only from the compact star forming knots where the star clusters are located. Using larger apertures in knots B and C would result in higher and lower temperatures in these knots, and consequently in lower and higher metallicities respectively. Knot A is scarcely affected because its temperatures and metallicities are similar to the values of the surrounding gas.

\subsection{Nitrogen abundances}\label{section:nitrogen_abund}

\begin{figure*}
    \centering
    \includegraphics[width=2\columnwidth]{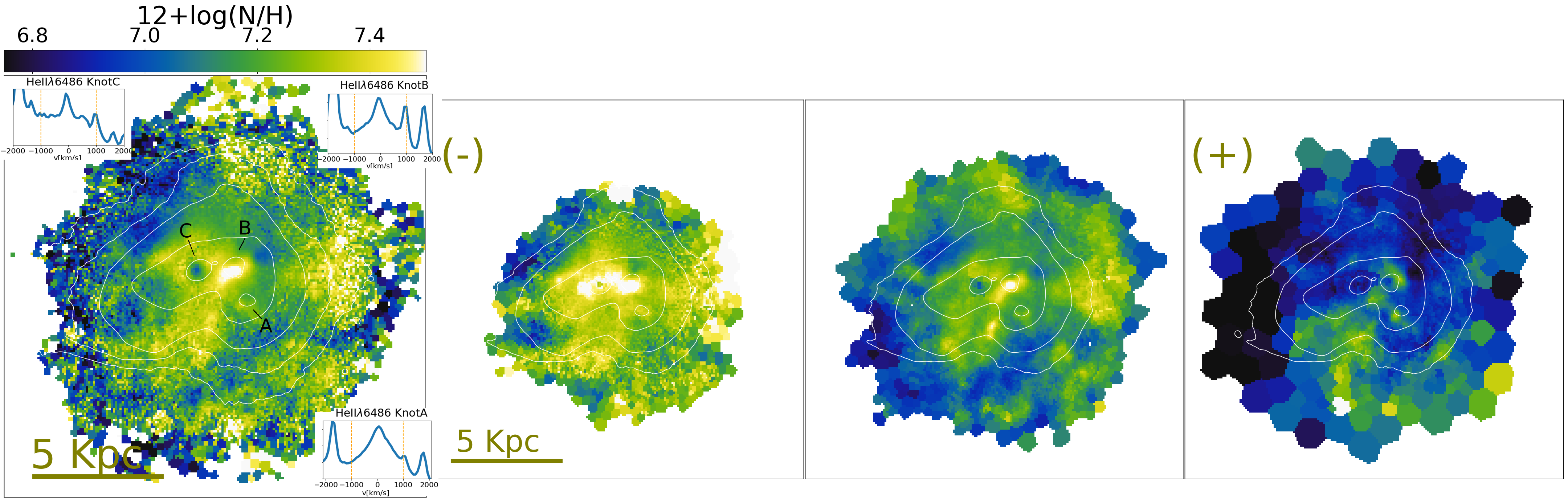}
    \caption{Nitrogen metallicities derived from the direct method. The top panel shows the integrated metallicities while the three smaller panels in the second row display the metallicities of the blueshifted (-), central (c) and redshifted (+) gas. For the three knots we show insets of the HeII$\lambda$4686 spectral line, which is an indicator of the strong Wolf-Rayet winds.}
    \label{fig:element_abundance_N}
\end{figure*}

\begin{figure*}
    \centering
    \includegraphics[width=2\columnwidth]{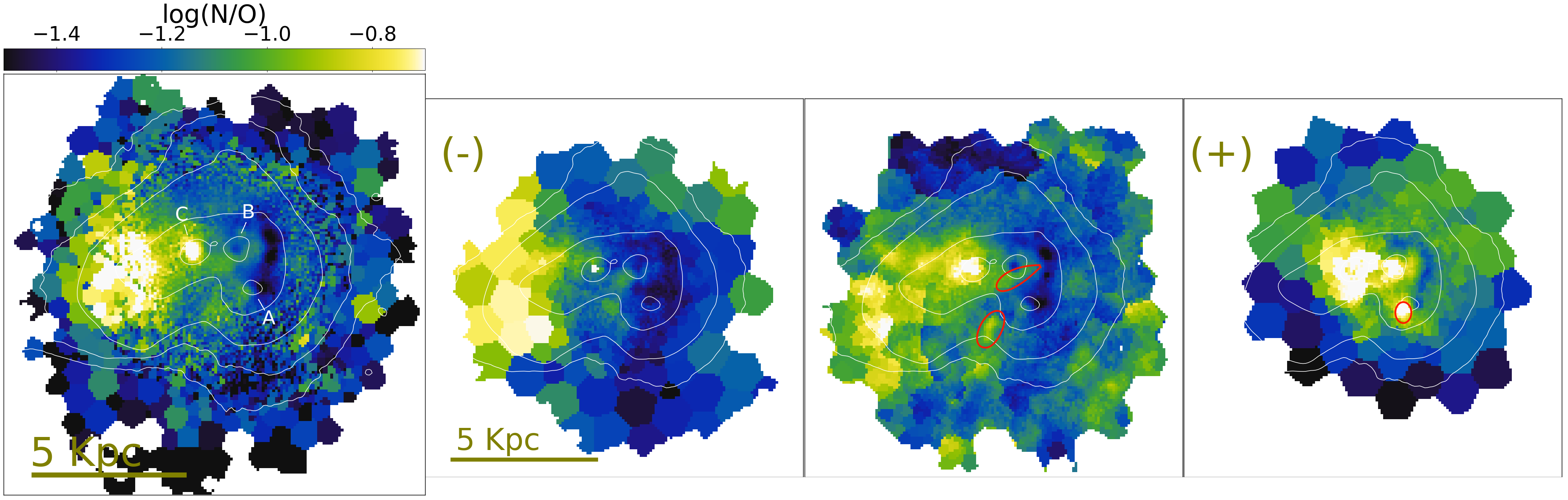}
    \caption{Nitrogen-to-Oxygen abundance ratio. The top panel shows the integrated metallicities while the three smaller panels in the second row display the metallicities of the blueshifted (-), central (c) and redshifted (+) gas. In red we mark the places with enhanced nitrogen abundances attributed to Wolf-Rayet stars}
    \label{fig:N_over_O}
\end{figure*}

In Fig. \ref{fig:element_abundance_N} we present the nitrogen metallicity maps up to a distance of 7.5 kpc in radius. The high intensity of the [N II] lines allow us to extract high-spatial resolution information with unprecedented details from the halo. The distribution of the zones with the highest nitrogen abundances are somewhat different relative to the oxygen abundances. The nitrogen-rich zones extend more towards the south of knot C and after $\sim$5 kpc bend to the west. Thus, the nitrogen-enriched gas seems to stream from knot C towards the southeast, reaching and enhancing part of the shocked area that is considerably poor in oxygen elements. Another noticeable difference is found knot B, that is nitrogen-enriched in an elongated-fashion rather than a point source. Despite these differences, both metals are likewise enhanced in the highly ionized cones and the filamentary structure in which the gas is mostly populated by doubly ionized ions (accounted here through the ICF).

The nitrogen abundances from the gas at different velocities show, similar to oxygen metallicities, a decreasing enrichment from the blueshifted (12+log(N/H) $=$ 7.42$\pm$0.05) to redshifted gas, with an average drop of 0.2$\pm$0.04 dex between the channel maps.
In the knots the nitrogen abundances vary significantly. Knot A has intermediate values (7.28$\pm$0.02) while knot B and C have the highest (7.64$\pm$0.02) and lowest (7.01$\pm$0.02) values respectively. 
In the left panel of the nitrogen abundance map (Fig \ref{fig:element_abundance_N}), we additionally show three insets displaying the profiles of the HeII$\lambda$4686 line from the star forming knots. The broadness of this line (v$\geq$1000 \kms\ ) is commonly use as a direct tracer of the Wolf-Rayet (WR) fast winds. After the intermediate-mass stars, these winds are a second source of nitrogen-rich gas and can be traced in regions around Wolf-Rayet clusters. 

\citet{James2013} analysed the Wolf-Rayet population with dominant ionized nitrogen lines (type WN) in Haro 11, and estimated a total population of 5200$\pm$1500 WN stars. In the knots, the authors estimated a total amount of 2700$\pm$650 WN stars, most of them located in knot B, followed by knot A and in lesser quantity by knot C. The remaining WN stars then might be located outside the knots, a large fraction of them perhaps in the star forming regions of the dusty arm where we observe also a broad WR feature around 4680 \AA . 
In knot A, we find that the western part, that probably belongs to the dusty arm, has the brightest and broadest He II$\lambda$4686 emission in the galaxy, but it is not nitrogen enhanced compared to its surroundings (12+log(N/H) $=$ 7.10$\pm$0.04). In the southern part of this knot both conditions are true; we observe broad He II lines and a nitrogen-enhancement of 0.24$\pm$0.03 at redshifted velocities. 
Knot B, who shows also a broad HeII emission, is locally enhanced in an elongated form at central velocities (12+log(N/H) $=$ 7.60$\pm$0.03). Both knots A and B host several massive star clusters at ages where Wolf-Rayet stars are active \citep[$<$ 5 Myr]{Adamo2010}, but clearly not all Wolf-Rayet clusters surroundings shows local enhancements of nitrogen. 
A second elongated structure that is nitrogen-enriched by 0.23$\pm$0.04 relative to its surrounding is found southeast of knot A at central velocities (encircled with red in fig. \ref{fig:N_over_O}), but it has no broad HeII$\lambda$4686 emission. We searched for young massive star clusters in this region from the catalogue of \citet{Adamo2010} and found there a 7.5 Myr old star cluster, which is older than the typical lifetime of Wolf-Rayet stars, and a group of young massive clusters about 600 pc north of this feature, but also here without an associated broad HeII$\lambda$4686 emission. Thus, the presence or absence of WR signatures linked to nitrogen-enhancements suggest different phases of the evolutionary path of WR stars and their impact on the nitrogen enhancements of their surroundings. The different scenarios found here will be discussed in the next section.

\subsection{N/O abundance ratios}\label{section:n_over_o}

At low metallicities, it is thought that nitrogen is primarily produced in the CNO cycle of massive stars. Additional factors such as the population of Wolf-Rayet stars (WR) and stellar rotation play an additional role in the nitrogen enrichment of metal-poor environments, hence in this metallicity regime the N/O relation is not well established. At higher metallicities, the nitrogen is mainly produced in intermediate-mass stars and the N/O ratio steeply increases. An effect of the metal-enrichment by different stellar populations is the delayed ($\geq$100 Myr) nitrogen-enrichment with respect to the the onset of star formation, because intermediate-mass stars (M $=$ 2-8 \Mo ) live considerably longer ($\geq$50 Myr) than massive stars where oxygen elements are produced. Thus, the nitrogen-to-oxygen abundance ratios provide important information about the stellar components of the progenitors and the evolutionary stage of a galaxy \citep{Garnett1990,Andrews2013,PerezMontero2009, Maiolino2019}. 

We present the log(N/O) maps in Fig. \ref{fig:N_over_O}. 
The distribution of the highest ratios in the map (avg. log(N/O) $=$ $-$0.69$\pm$0.7), suggest that, the nitrogen-enriched gas originates in knot C and streams to a wide region towards the eastern halo that belongs to the shocked area, while mixing with the halo gas. 
Although Knot C has, in general, low metallicities, its gas is considerably more nitrogen-enriched for its oxygen abundance ($-$0.63$\pm$0.02) than any other star forming region in the galaxy. Knot C hosts many massive star clusters in average older than the stellar populations in knots A and B. Thus, besides the previous and current Wolf-Rayet population, intermediate-mass stars from older star clusters might have started to boost considerably the ISM with nitrogen. Because of the low densities in this knot, metal-rich winds and outflows might move outwards unhindered, enriching the halo with metals. 

Knots A and B have different gas conditions and metallicities than knot C. Knot B has a broad HeII$\lambda$4686 line and an obvious nitrogen-enhanced elongated structure at central velocities (marked in red in Fig. \ref{fig:N_over_O}), but the highest log(N/O) ratio of $-$0.59$\pm$0.05 at redshifted velocities. The kinematics of the \NIIr\ line have shown nitrogen-rich outflows with redshifted velocities in this knot. The combined information suggests that we are tracing the nitrogen-rich winds from Wolf-Rayet stars blowing in the opposite direction. In knot A we find two scenarios. In a group of massive star clusters located towards the west of the knot, we observe broad HeII$\lambda$4686 lines but normal, non-enhanced log(N/O) ratios (avg. $-$1.42$\pm$0.3). Towards the south of knot A, there is another group of young massive star cluster with broad HeII$\lambda$4686 emission and log(N/O) ratios that are 0.3 dex higher at redshifted velocities than the knots surrounding. Finally, the region of the second feature southern of knot A (marked in red in Fig. \ref{fig:N_over_O}) displays also relatively high log(N/O) values (avg. $-$0.86$\pm$0.4), but no broad HeII$\lambda$4686 emission. Only a 7.5 Myr star cluster was identified here.

Some studies in other starburst galaxies have shown that broad HeII$\lambda$4686 emission and enhanced nitrogen do not always occur together. In NGC 5253  \citep{MonrealIbero2012} and ESO 338-IG04 \citep{MonrealIbero2012, Bik2018} similar scenarios were found as described above and were linked to nitrogen-rich wind material from Wolf-Rayet stars and the posterior dissemination of this gas in the ISM. 
Thus, we might witness a different evolutionary stage of Wolf-Rayet clusters. The absence of nitrogen-enhancements in a region with strong He II signature suggests at a very young Wolf-Rayet population that do not have had the time to enrich with nitrogen the surrounding medium. 
Regions with He II emission and nitrogen enhancements are associated with massive Wolf-Rayet clusters at intermediate age. Knot B and partially knot A are embedded in a dense nebula, so that the nitrogen-rich winds of these stars might have not dispersed, mixed and diluted yet. A case of nitrogen enhancements but no  HeII$\lambda$4686 emission was found in NGC 5253 and is attributed to a delay in the incorporation of the nitrogen-rich material into the ISM, which can take few Myrs after the lifetime of Wolf-Rayet stars \citep{MonrealIbero2012}. 

The log (N/O) values from the entire galaxy of $-$1.17$\pm$0.01 is somewhat high in relation to the galaxy oxygen abundance. It lies above the expected values for nitrogen to be produced only from massive stars in the fit of \citet{Andrews2013}; but in the reference line for the nitrogen abundances to be partially enhanced by intermediate mass stars in the chemical evolution models from \citet{Vincenzo2016}.

\subsection{Sulphur abundances}

We present the sulphur metallicity maps in Fig \ref{fig:element_abundance_S} in the appendix. Sulphur, like oxygen, is an $\alpha$-element produced mainly in massive stars and released out into the ISM through core-collapse supernovae, hence their element abundances are likewise distributed. 
Comparing the sulphur and oxygen abundances at central velocities, the former seems to be less enriched around knot B and towards the southern filaments. 
Indeed, using a correlation derived by \citet{Milingo2010}, we derive about 0.2$\pm$0.6 dex lower sulphur abundances around knot B than the values expected from HII-regions of similar oxygen abundances. These authors found a tight correlation ($12+log(S/H) = -1.63 + 1.01 * (12+log(O/H)$) for HII-regions but an average offset of 0.45 dex for Planetary Nebulae (PNe), which tend to have lower sulphur abundances. 
\citet{Pottasch2006} suggest that this anomaly would be expected in C-rich environments and might result from the formation of some dust that tend to favour the formation of sulphides (e.g. MgS, FeS). 
The depletion of sulphur in dust grains is a long-standing problem in astrochemical models and observations.
Early works have shown that sulphur is depleted in dense environments, but remains atomic and close to the cosmic abundances in primitive ISM environments, or even in the diffuse phase of molecular clouds \citep[e.g.][]{Jenkins2009}. Current models predict that most of the sulphur is depleted in some kind of S-S bounded molecules, especially H$_2$S, however considerably lower amounts than expected were found from observations. Despite that new gas-grain models are reproducing the depletion of sulphur in a mix of sulphur-bearing species on grains across elements like oxygen, carbon, nitrogen and hydrogen \citep[][]{Laas2019}, the sulphur depletion problem is still unsolved. In the case of Haro 11, dust might play a role in lowering the sulphur abundances given that the highest discrepancies are found around knot B, which is more dusty.

\subsection{Metallicities derived from the S strong-line method}

We additionally derive the abundances of Haro 11 from a strong-line method that is temperature-independent. We follow the "S-calibration" (S-calib) presented in \citet{Pilyugin2016} that use the strongest forbidden lines in the optical spectra relative to \Hb . These are: \OIIIHb , \NIIHb\ and \SIIHb . Among many strong line diagnostics, we chose the S-calibration from \citet{Pilyugin2016} because it minimises the effect of the ionization parameter and was found to have a low scatter ($<$0.2 dex) in a large metallicity range with respect to the temperature-dependent method.  This metallicity prescription is calibrated based on HII regions in which the abundances were previously derived using the direct method \citep{Pilyugin2016, Maiolino2019}. 

\begin{figure*}
    \centering
    \includegraphics[width=2\columnwidth]{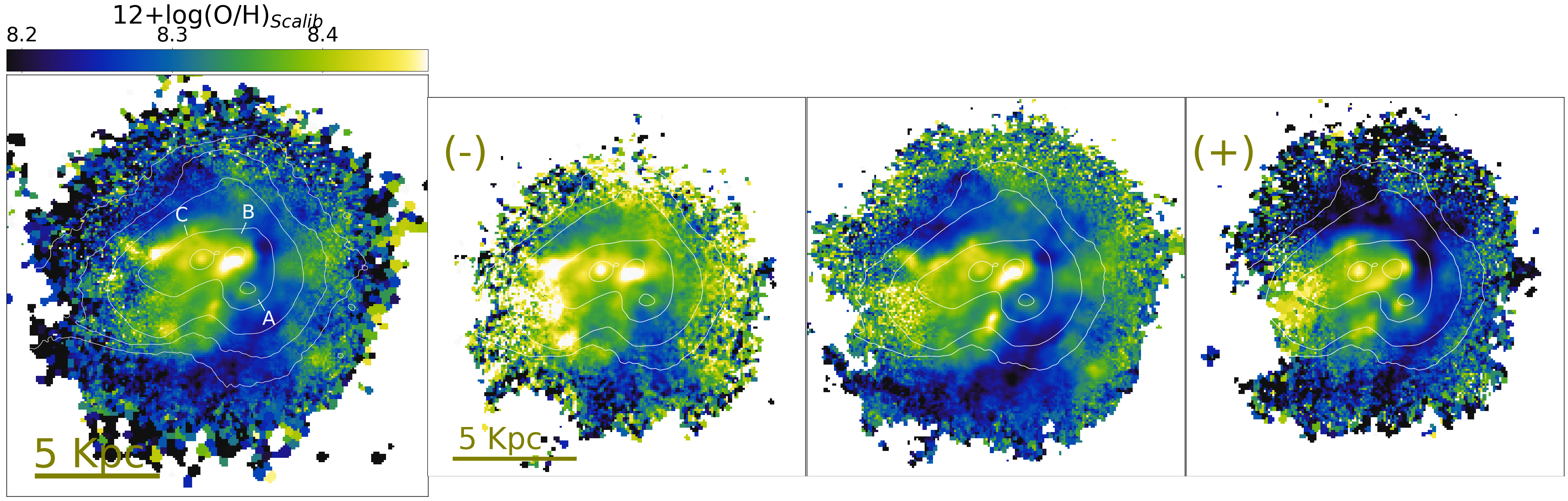}
    \caption{Oxygen abundances derived from the strong-line method using the S-calibration of \citet{Pilyugin2016}. }
    \label{fig:element_Scalib}
\end{figure*}

Fig. \ref{fig:element_Scalib} shows the resulting oxygen metallicities up to a distance of 7.5 kpc in radius. The estimated values are in general low, with an average value of 12+log(O/H) $=$ 8.28, Std Dev $=$ 0.013. 
Comparing these with the metallicities derived from the direct method (Fig. \ref{sec.abundances}), both are substantially different, not only in the range of metallicities, but also in how the metal-enhanced regions are distributed. The same metallicity range was only obtained in the direct method after using a 10$^4$ K constant temperature instead of our estimated temperatures, however, the discrepancy still persists in the distribution of the metal-enhanced regions. 
In the strong-line method, these metal-rich regions strongly correlate with the regions showing the highest nitrogen enhancements. Moreover, both elongated nitrogen-rich features tracing Wolf-Rayet outflows are clearly recognisable at central velocities. This effect might arise from these regions being highly nitrogen-rich, element that is not accounted for in the oxygen metallicities from the direct method, but of this particular strong-line method calibration. 

Here the highest metallicities seem to trace the path of a streaming metal-enriched gas originated in knot C. These regions are not metal-enhanced in the oxygen abundances from the direct method, especially in the shocked area and knot C. Moreover, both methods give different results in the highly ionized areas: the ionized cones and the filamentary structure.
Notwithstanding, both methods roughly agree in the metal-enhancements of knots A and B, and the western ionized cone. Comparing the metallicities at different velocities, the blueshifted gas is more metal-enriched than the gas at central and redshifted velocities. Similar trend was found in the direct method and confirm the enhanced oxygen-abundances in the approaching gas, independent of the method that is used. In the section \ref{subsec:temp and abund variations} we will discuss in more detail the reliability of the direct method and the strong-line method in the regions where we found the largest discrepancies.

\section{Discussion}

\subsection{Where do the nitrogen and oxygen enrichment come from?}\label{sec:SN_knots}

We analyse the cluster population to get insight into the sources of recent metal enrichment in the galaxy. We use the catalogue of the star clusters identified by \citet{Adamo2010}, except for knot C whose stellar mass was considerably underestimated\footnote{\citet{Adamo2010} underestimate the total fluxes of knot C due to their assumption of knot C being a point source \citep{ostlin2015}}. Knot C is likely a nuclear star cluster (M$_{vir}\sim 10^{8.4}$, r$_{eff}=$40 pc, \citealp{ostlin2015}) with ongoing star formation for perhaps hundreds of Myrs \citep{Neumayer2020}, though the last $\sim$40 Myrs in a star bursty episode. Here we analyse only the clusters that we can directly link to the metal-enrichment of their surrounding gas (Age$<$200 Mys). Because of the merger dynamics, older clusters might have considerably migrated from their formation places (we counted 6 massive older clusters at ages from 0.5 to 14 Gyrs spread out in the halo at r$<$5kpc).

Oxygen is released shortly after the onset of star formation when the most massive stars commence to explode as supernova. To understand better the intensity of these events in the knots, we run for each cluster-mass the evolutionary synthesis code \texttt{STARBURST 99} \citep{Leitherer1999} with a Kroupa IMF. For all clusters but knot C we set the run for an instantaneous star formation rate. For knot C we set a continuous star formation rate of 4 \Mo\ yr$^{-1}$, which is equivalent to a total stellar mass of 10$^{8.2}$ \Mo\ built up in 40 Myrs. Each run provides the cluster supernova rate in time-steps of 0.1 Myrs that can be easily converted into the number of supernova events in a specific time of the cluster evolution. We find that in the last 10 Myrs 81\% of all supernovae happened in knot C. From the remaining supernova events, only 7\% happened in knot B while about 10\% in Knot A.

From this result we might expect knot C having the highest oxygen abundances from the star forming knots, however the oxygen abundances derived from both methods are considerably low. The nitrogen-to-oxygen abundance ratio on the other hand is very high and can be explained by the continuous star formation over the past hundreds of Myrs \citep{Adamo2010} so that both, previous and current Wolf-Rayet clusters and intermediate mass stars in older clusters have contributed or are still contributing to the nitrogen-enrichment in the ISM of knot C. Despite this, the nitrogen abundances of knot C are the lowest from the knots. 
Thus there is a significant deficit of metals in knot C, especially $\alpha$-elements. 
One possible explanation for the low oxygen abundances could be due to accretion of a significant amount of metal-poor gas, induced perhaps by the merger dynamics, but, because of the low densities in the knot and its surroundings, the amount of this infalling gas have to be significantly low. A more plausible explanation could be that a great fraction of the oxygen-rich gas was expelled unhindered to the halo by the large numbers of supernovae in the knots, while the nitrogen-enrichment started more recently there. Similar explanation was favoured by \citet{James2013}. 
From ionization mappings in velocity bins, \citet{menacho2019} found evidence of a fragmented superbubble shell around knot C (r $\sim$ 1.7 kpc), which is supported by the low density cavity and high temperature found in this region. Thus, a great part of the supernova ejecta could have escaped, enriching the surrounding gas up to the halo or even out of the galaxy. From the number of supernovae, this knot is likely the principal source of recent metal-enrichment in the galaxy.

In contrast, knot B has mostly a very young and massive star cluster population where the SNe activity might have recently started. Owing to the concentrated oxygen-enhancement in the core of the knot, a considerable fraction of the recent oxygen-enriched SN ejecta seems to not have had the time to propagate and mix with the halo gas. 
Additionally, knot B is nitrogen-enriched in an elongated-fashion by the winds of the Wolf-Rayet population, as seen in other galaxies like ESO 338-IG04 \citep{Bik2018}. Because of the young population of knot B, intermediate mass stars are likely not playing a role in the nitrogen enrichment of this knot. 

In knot A, the star cluster population is predominantly younger, but it hosts also several massive clusters older than 10 Myrs. 
Although in the past Myrs a large number of supernovae exploded in knot A, its ISM is oxygen-enriched only in part of it. Moreover, in the non-enriched area, the metallicities are similar to its closest surrounding, which is slightly lower than the galaxy average metallicity, despite that about half of the massive cluster population producing supernovae in knot A, are located here. 
In knot A we also measure a relatively low nitrogen-abundance, notwithstanding the high Wolf-Rayet population evidenced by the strong HeII$\lambda$4686 feature, and the likely significative number of intermediate-mass stars from the older cluster population.
Moreover, the highly ionized zones of the southern halo are nitrogen-underabundant with respect to their surrounding. This metal-enriched gas could be hot and escaping out of the galaxy through the low density channels towards the south, while enriching the IGM with metals.

\subsection{Temperature and metallicity discrepancies}\label{subsec:temp and abund variations}

\begin{figure}
    \centering
    \includegraphics[width=\columnwidth]{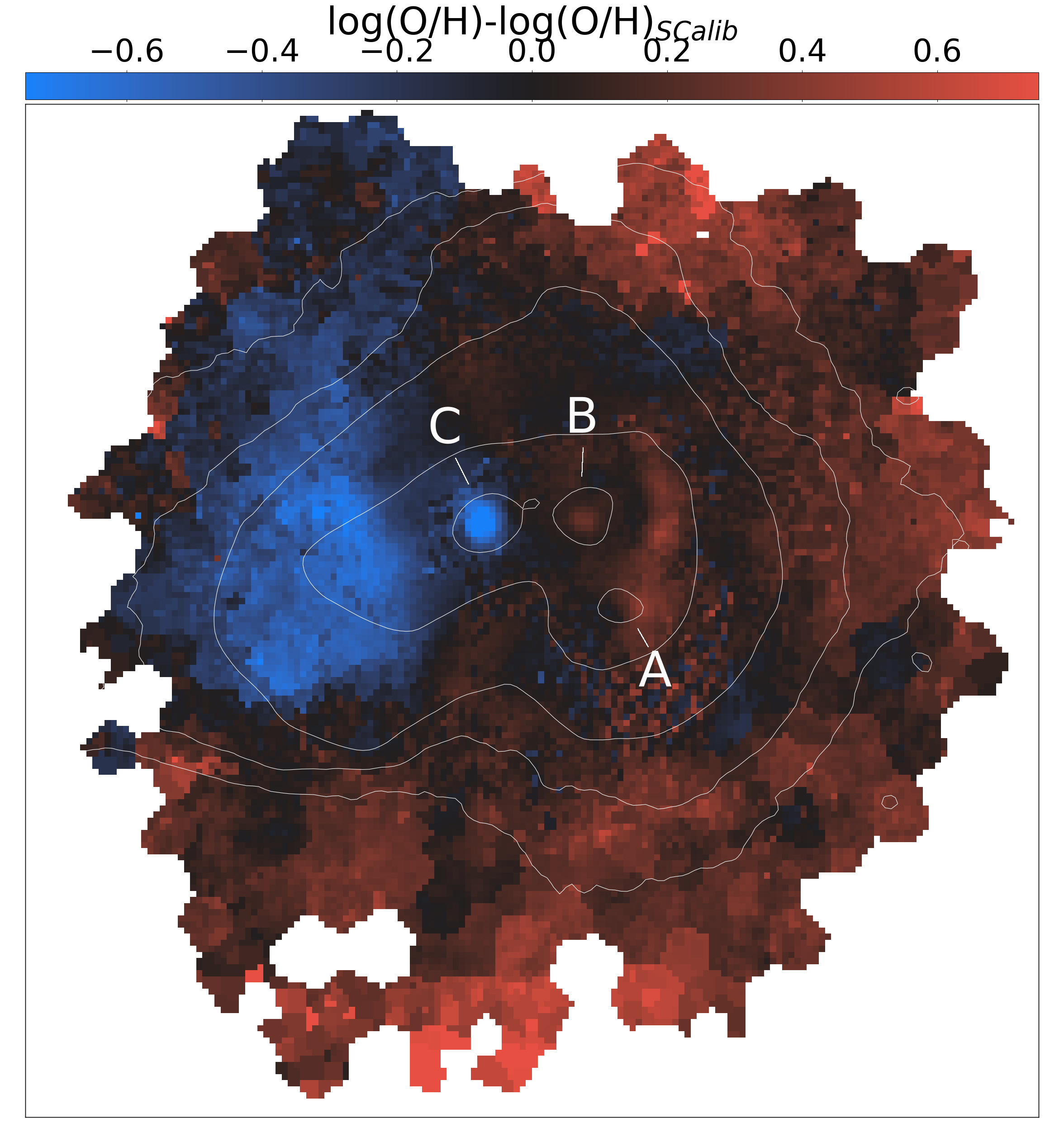}
    \caption{Oxygen abundance discrepancy between the abundances derived from direct method and the  strong-line method. This variation is as high as 1 dex in knot C, the shocked zone and towards the south and north where the highly ionized filaments and the ionized cone.}
    \label{fig:abund_direct_vs_strong_method}
\end{figure}

We have previously shown that the metallicities obtained from different techniques differ considerably in some regions of Haro 11. Fig. \ref{fig:abund_direct_vs_strong_method} shows the differences of the oxygen abundances derived from the direct method, which has a tight dependence on the temperature, and the strong-line method which uses strong line ratios only. In most parts of the galaxy we derive higher abundances using the direct method, especially in the dusty arm and in the highly ionized regions where this discrepancy is larger than 0.3$\pm$0.085 dex. Contrarily, in the shocked area and knot C the abundances derived from the strong line methods are more than 0.5 dex higher. So, which are the real abundances in Haro 11? Which method provides the most reliable results? The temperature-dependent method is typically favoured because it directly measures the element abundances from simple relations that are drawn from the atomic physics. On the other hand, strong line methods are secondary calibrated techniques that can be degenerated with respect to some properties of the nebula like pressure and ionization \citep{Maiolino2019}. Here we present some problems that could have affected our abundance derivation in both methods.

One of the greatest discrepancies (up to 0.6 dex) is found east of knot C in the largest neutral and lowly ionized gas region (7$\times$3 kpc$^2$) that seems to be hit by fast shocks \citep{menacho2019}. This area, likely compacted by the merger dynamics, is located close to knot C, which in the last 10 Myrs hosted most of the supernova events from the central starburst region. High cadence of SNe created recurrent shocks that not only seems to have induced a high turbulence in this area ($\sigma$ up to 100 \kms ), but they might have affected the emission lines. Although it is not very well understood how shocks operate in the ISM of galaxies, it has been shown that shocks can alter the properties of the emission lines in such a way, that these deviate from the condition of the gas where models and relations are derived. Firstly, in the shock fronts the gas deviate from a Maxwellian distribution \citep{McKee1980}, a condition that is needed for the estimations of the temperature; and lastly, shocks can enhance the strength of the collisionally excited lines, in particular the auroral lines, which leads to an overestimation of the temperatures \citep{Peimbert1991}. This effect might explain the highest temperatures (up to 15000 K) and consequently low abundances estimated in this area. If we assume a constant temperature of 10$^4$ K, we derive an oxygen abundance of 12+log(O/H)$=$8.3 ($\pm$0.05) with the direct method, which is somewhat similar to the value of 8.39$\pm$0.02 derived for this region from the strong line method. 
Shocks also affect the nebular lines, although weakly, that are used to derive other physical properties like the density and ionization. Here we can not estimate quantitatively the effect of these shocks in our density estimations.

The highest abundance discrepancy of about 0.79$\pm$0.03 dex is measured in knot C. In this knot we also obtained the highest [S III] -- [N II]-temperature discrepancy, despite that those ions populate the same gas. For the temperature calculations, we estimate indirectly the effect of recombination excitation in enhancing the [N II]-auroral line and found an excess of about 50 per cent. But deriving this from the [O II]-auroral lines, that is even more affected \citep{Rubin1986}, the excess was estimated to be $<$5 per cent. Knot C is a potent source of soft-X ray and FUV emission in the galaxy \citep{hayes2007,grimes2007} and the possible source of the observed LyC radiation \citep{Ostlin2021}. Additionally, there is some dust and molecular gas likely shielded in dense clouds feeding this star formation region. Thus, given the hard ionization field and the high SN rate, a large range of energy transport processes like plasma waves, shocks, ion heating, fast electrons produced by photoionization with X-ray or extreme UV photons might take place in this knot. Such environments favour a $\kappa$-distribution of electrons that is a Lorenzian-like distribution \citep{Nicholls2012,Dopita2013}. One effect of the $\kappa$-distribution is that the collisionally excited lines are weak, but the Balmer lines strong. This effect is observed in knot C and can explain the low \OIIIHa\ ratios, that were used to trace the ionization state of the gas in \citet{menacho2019}, or even the low metallicities in the direct method because they are derived from metal-line relative to \Hb . In such condition not only the temperature and hence the abundances from the direct method will be affected, but also the abundances derived from the strong-line method. \citet{Ferland2016} note however, that the time needed to thermalise these electrons is shorter than the heating/cooling timescales, therefore this mechanism can be ruled out in most HII-regions and Planetary Nebulae. Other mechanisms like shocks, chemical inhomogenities, real temperature fluctuations, conduction fronts, etc. can affect the temperature measurements \citep{Peimbert1967,Peimbert1969,Peimbert1991, Esteban2002,Dopita2013, Nicholls2012}, however they might affect all temperatures derived from the auroral lines. Thus, we do not find a clear reason for the very large [N II]-temperatures ($\sim$52700 K) in knot C. It seems that, there is at least one mechanism that strongly enhances the \NIIt\ auroral line only. 
Besides, the metallicities seems to be likewise strongly affected and we are not confident with the results from either the direct method or the strong line method. 
From the integrated spectra of the entire galaxy, the [N II]-temperature is about 30000 K higher than the [O I] and [S III]-temperatures. This trend was observed in other Lyman break Analogs \citep{Loaiza2019} and might indicate extreme conditions in some regions within those galaxies. 

On the other hand, the direct method yields higher abundances in the highly ionized regions compared to the strong line method. At first glance, this effect could arise from inaccurate temperature estimations, due to a weak [S III]-auroral line in the farthest cells (despite the signal-to-noise of at least 6), however this tendency is observed in all the filamentary structure, regardless of the signal-to-noise ratio. On the other side, strong line methods are known to be degenerate with the ionization parameter and deliver imprecise values in the high and low abundance ends \citep{Dopita2013}. 
Thus, we can not find a straightforward explanation for this discrepancy. 
Similar issues in the strong line diagnostics were found in high resolution data from recent works that were probing the metal abundance of the gas on spatial scales smaller than typical HII regions \citep[e.g.][]{James2016}. 

It seems that traditional relations developed from averaged measurements of the line strengths or simplified method, fail to probe the diverse conditions of the gas in extreme environments. 
Despite that we are probably averaging out to some extent the gas properties from more than one kinematic component (a lesser issue in the channel maps), the discrepancies found in the gas under extreme environments are evident. Such extreme condition might be ubiquitous in galaxies with extraordinary star formation, condition that is more frequently observed in the high-redshifted galaxy population. 
Moreover, the new generation of instruments (like JWST or ELT) will allow us to study the ISM of galaxies at high spatial and spectral resolution, so that any discrepancies that were averaged out 
in the low spatial/spectral resolution data will become evident.
Thus, we are in the need of robust relations that include realistic models where several physical processes are simultaneously at work.

\subsection{The superbubble structure. Evidence of the supershell?}\label{subsection:superbubble}

To verify if the arc-like feature, drawn from the [N II]-most-blueshifted gas relative to the \Ha\ gas in the ${\Delta}v_{(H\alpha-[N II])}$ map (Fig. \ref{fig:velocityHaNII}, and the bright \Ha\ arc are delineating the same structure, we superimpose this as contours in the \Ha\ map and in the velocity dispersion map (Fig. \ref{fig:shell_vel}).  We find that both semi-arc fit well. Even more, together they seems to delineate a circle of about 3.8 kpc in radius that cover the entire starburst region. If so, both structures seems to be the shell-remnants of the superbubble suggested by \citet{menacho2019}, whose breakout might have created the large filaments in the southern hemisphere. 
In most part of this shell, the velocity dispersion is slightly greater than 100\kms\ , larger than in its surrounding gas. Supernovae-induced shocks are likely hitting this structure while inducing turbulence on it.

\section{Conclusions}

In this paper we take advantage of the spectro-photometric capabilities of MUSE to map the physical properties of the ionized gas of Haro 11, the closest confirmed Lyman continuum leaking galaxy. 
Haro 11 is a merger system with a vast population of predominantly young massive star clusters, whose radiative and mechanical feedback has a great impact not only in the dynamics of the system, but in the properties of the gas itself.
The merger dynamics and the strong stellar feedback have imprinted the halo with discrete structures, most of which are traced solely in the gas at blueshifted, central or redshifted velocities. 
Here we study the integrated properties of the gas as well as the properties of gas at blueshifted, central and redshifted velocities.
Our deep observations allow us for the first time to map the kinematics, dust extinction, density, temperature and metallicity of the gas to farther distance in the halo. 
In summary:

\begin{figure}
    \centering
    \includegraphics[width=\columnwidth]{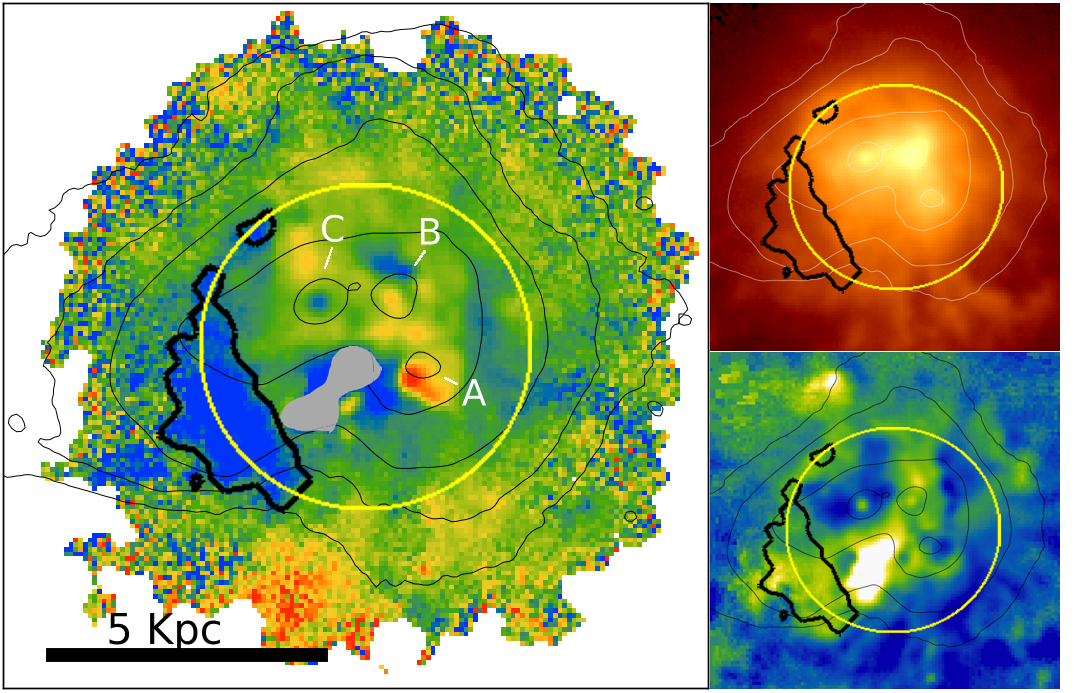}
    \caption{Large panel, velocity difference of the \Ha\ and \NIIr\ gas (same as in Fig. \ref{fig:velocityHaNII}. Right panels, the \Ha\ map at blueshifted velocities (Fig. \ref{fig:HaFlux}) and the velocity dispersion map (Fig. \ref{fig:velocity}). In black contours we show the areas with [N II]-blueshifted velocities larger than 30 \kms . The circle shows the apparent size of a shell that both structures seems to trace. This shell of about 3.8 kpc in radius seems to trace the remnant of the superbubble suggested by \citet{menacho2019}, whose breakout formed the $\sim$10 kpc large filaments in the southern hemisphere. 
    }
    \label{fig:shell_vel}
\end{figure}

\begin{itemize}
    \item {The \Ha\ gas shows complex velocities in both, the central starburst and the halo. 
    We trace a velocity gradient in the central part of the galaxy suggesting virial motions of both galaxy progenitor cores, knots B and C, in process of coalescence. 
    The velocities in the halo are on the northern side dominated by virial motions of one galaxy progenitor, on the southern side however by a filamentary structure that was created by momentum feedback. 
    This suggest that both, the merger dynamics and mechanical feedback from supernovae and stellar winds are equally impacting the global kinematics of Haro 11. 
    }
    \item {By comparing the [N II] and \Ha\ velocities we trace an arc structure from the [N II] gas that has relatively high blueshifted velocities with respect to the \Ha\ gas.
    This structure connects well to a northern turbulent arc seen in the \Ha\ and $\sigma$ maps and together form a semi-circle of 3.8 kpc in radius. 
    Both structures are perhaps remnants of a kpc-scale superbubble, whose break out might have created the filamentary structure towards the south as suggested by \citet{menacho2019}
    }
    \item {We measure high velocity dispersion in areas surrounding knot C at a radius $\geq$ 1.5 kpc and the northern halo. This is not surprising since in the last 10 Myrs about 80 per cent of the supernova events happened in knot C. 
    This high-cadence of supernova explosion propagates recurrent shocks that are clearly inducing high turbulence to its surrounding gas. Two structures are especially affected by these events: 
    the \Ha\ arc towards the north where we measured high turbulence, and a large lowly ionized structure towards the east where fast shocks were detected \citep{menacho2019}.
    }    
    \item {The extinction map suggests low dust content in the halo, but moderate in the knots and the dusty arm. The dusty arm is composed by four dusty lobes, each with distinct characteristics. 
    In knot B we uncover a dusty bubble with a dust-free interior of $\sim$430 pc likely created by radiative feedback from the young and massive cluster population at knot B. 
    We additionally found evidence of dusty gas inflows in the area of knot B from the analysis of the intrinsic extinctions in the blueshifted, systemic and redshifted emitting gas.
    }
    \item {We derive relatively low densities in the entire galaxy though moderately high 
    in knots B and A. In knot C and around it we measure the lowest values of the galaxy. This support the evidences seeing in the ionization mapping \citep{menacho2019} of an fragmented and evacuated bubble around knot C, created by its own massive stellar population. 
    }
    \item {We determine the electron temperatures from the [S III] auroral and nebular lines.
    The temperature map is highly variable in the halo and the knots. High temperatures ($\geq$15000 K) were measured in knot C and in areas of the halo where fast shocks were previously detected \citep{menacho2019}. 
    In the latter the temperatures are likely altered by effect of shocks that tend to enhance the auroral lines and hence overestimate the temperature estimations \citep{Peimbert1991}.
    }
    \item {We additionally derive the temperatures from the [N II] lines, which arise from a gas with the same ionization as the [SIII] lines, and from the [O I] lines that trace a more neutral gas. The [O I] and [S III]-temperatures agree well overall, however the [N II]-temperatures are considerably higher. A particular case is given in knot C where we measure extremely high [N II]-temperature of $\sim$ 45000 K. [N II]-Temperature discrepancies were found in several starbursting galaxies  \citep{Garnett92, Loaiza2019} and might indicate extreme conditions in some regions of the galaxy. Although it is still unknown what causes it, our data suggest that there is at least one mechanism that enhances the [N II]-auroral line only.
    }
    \item {We derive the metallicity of Haro 11 using a temperature-dependent method (direct method) and a strong-line method calibrated by \citet{Pilyugin2016}. We find large discrepancies in the results. Part of these discrepancies originate from the effect of shocks in overestimating the temperatures, other arise in the highly ionized zones, but the largest discrepancy is found in knot C. Excluding the shocked area, we do not find a clear reason for such diverse results. 
    }
    \item {
    We find that traditional relations derived from averaged measurements of line strengths or simplified methods, fail to prove the diverse condition of the gas in extreme environments, like in knot C. 
    Such discrepancies will become more evident in the era of JWST or ELT because of their high spatial/spectral resolution that will highlight regions with any discrepancies that were previously averaged out in the low spatial/spectral resolution data.
    }
    \item {Despite the diverse results in the metallicity distribution, both methods suggest however that the blueshifted gas is considerably more metal-enriched than that of the redshifted gas. We find that the different metallicities can be partially attributed  to the different metallicities of the galaxy progenitors; and partially to metal-enriched outflows from supernova ejecta that seems to flow preferentially toward us.
    }
    \item {The metallicity in the knots differs considerably. Despite the low values measured in knot C, this knot is likely the main source of the metal enrichment in the halo. In Knot B, the metallicities are strongly enhanced in the core of the knot suggesting at recent metal-enriched SN ejecta that did not have the time to propagate and mix with the halo gas. In knot A however, the metal abundances are comparable to its surrounding gas and we do not find evidence of its impact in the metal enrichment in the halo. The hot metal-rich gas from supernovae is perhaps escaping through low density channels towards the south.
    }
    \item {We uncover three nitrogen-rich redshifted outflows linked to Wolf-Rayet winds in knots A and B and one close to the doubly peaked broad region.
    }
\end{itemize}

\section*{Acknowledgements}

The authors acknowledge financial support from the Swedish Research Council and the Swedish National Space Board. 
This research has made use ASTROPY , a community-developed core PYTHON package
for Astronomy and APLPY \citep{Collaboration2013,Collaboration2018}, an open-source plotting package for PYTHON  \citep{Robitaille2012}.
This research has made use of NASA Astrophysics Data System Bibliographic Services (ADS). This study is based on observations made with ESO Telescopes at the La Silla Paranal Observatory under programme IDs 094.B-0944(A) and 096.B-0923(A).

\section*{Data availability}
The data underlying this article will be shared on reasonable request to the corresponding author.


\bibliographystyle{mnras}
\bibliography{Haro11_properties} 


\appendix
\section{Appendix}

\begin{figure*}
    \centering
    \includegraphics[width=2\columnwidth]{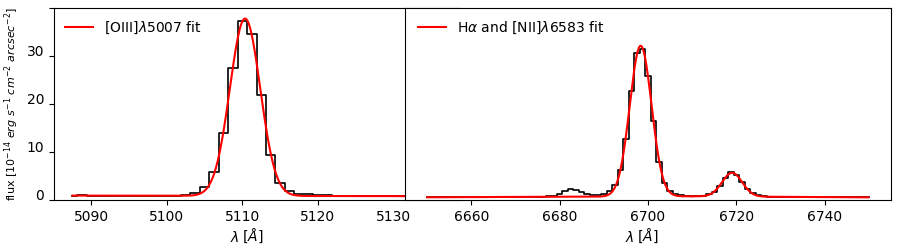}
    \caption{Gaussian fits to the \OIII , \Ha\ and \NIIr\ lines from the integrated spectra of the entire galaxy. The galaxy spectrum was taken from an aperture of 1\arcmin\ in diameter as described in sec. \ref{subsec:extractingSpectra}. 
    }
    \label{fig:fit_spectra_gal}
\end{figure*}

\begin{table*}
\normalsize
\centering
    \begin{tabular}{l|c|c|c|c|c|c}
    ~      & \Ha          & [O III]$\lambda$5007 & [N II]$\lambda$6583  & [O I]$\lambda$6300    & mean        & $\Delta$vel \\ \hline
    knot A & 6250.27$\pm$1.7 & 6250.52$\pm$0.9 & 6258.23$\pm$2.8 & 6251.88$\pm$12.2 & 6252.72$\pm$3.2 & 58.72$\pm$5.8    \\
    knot B & 6165.51$\pm$2.2 & 6174.12$\pm$1.4 & 6162.00$\pm$1.8 & 6178.04$\pm$6.7  & 6169.92$\pm$1.9 & -24.08$\pm$5.2   \\
    knot C & 6153.83$\pm$1.2 & 6147.92$\pm$0.8 & 6147.99$\pm$1.2 & 6178.70$\pm$3.8  & 6157.11$\pm$1.2 & -36.89$\pm$5.0  \\
    \end{tabular}
    \caption{Recessional velocities of knots A, B and C, estimated from the brightest lines from their 1\arcsec apertures. The fifth column shows the mean recessional velocities and in the six (${\Delta}vel$) we show the velocity difference of the different knots recessional velocities with respect to the central recessional velocity of the galaxy of 6194$\pm$4.9 \kms.}
    \label{tab:fit_vel_knots}
\end{table*}

\begin{figure*}
    \centering
    \includegraphics[width=10cm]{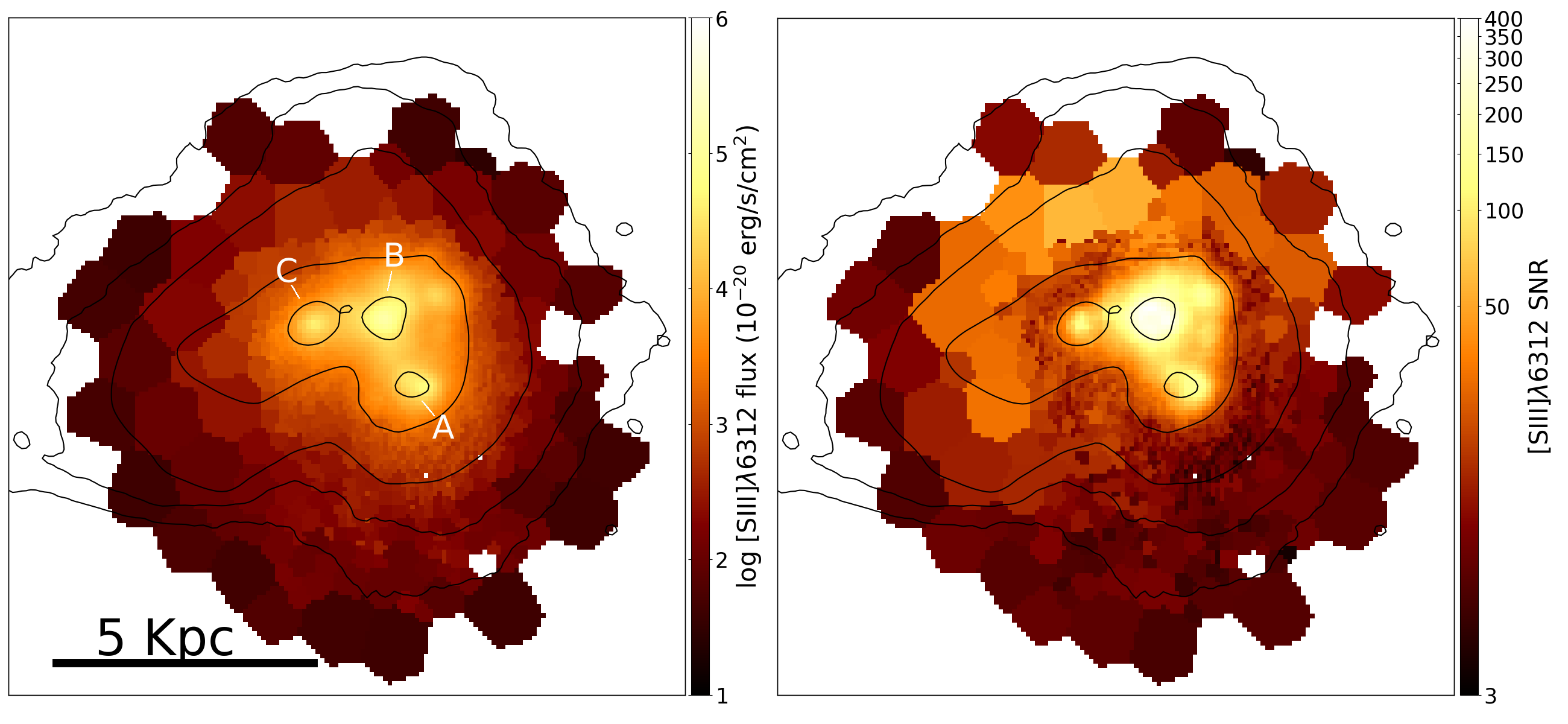} 
    \caption{Integrated flux and SNR from the stellar absorption corrected \SIIIb\ line. 
    }
    \label{fig:SIII6312_line}
\end{figure*}

\begin{table*}
\normalsize
\centering
    \begin{tabular}{lccc}
    ~                       & Centre of cavity$^{(*)}$ & cavity$^{(*)}$  & avg. value for dusty ring (shell) \\ \hline
    E(B-V)$_{blue}$             & 0.33                     & 0.37   & 0.29                             \\
    E(B-V)$_{cent}$             & 0.28                     & 0.37   & 0.46                             \\
    E(B-V)$_{red}$              & 0.55                      & 0.56   & 0.25                             \\ \hline
    E(B-V)$_{cent\_intr}$        & \textbf{- 0.04}                    & 0.01   & 0.17                              \\
    E(B-V)$_{cent\_intr\_INFLOW\_15\%}$ & 0.00                     & 0.06    & 0.21                              \\
    E(B-V)$_{red\_intr}$         & 0.27                     & 0.19   & \textbf{-0.03}                            \\
    E(B-V)$_{red\_intr\_INFLOW\_25\%}$ & 0.28                     & 0.33   & 0.09                             \\ \hline
    \multicolumn{3}{l}{$^{(*)}$\footnotesize{We refer to \textit{cavity} the dust-free interior of knot's B dusty bubble}}
    \end{tabular} 
    \caption{Extinction values of the dusty shell from knot B. The first column shows the extinction values of the very centre of the dust-free interior (\diameter =0.4 arcsec $\sim$170 pc), while the second column shows the average values of the whole dust-free shell interior (\diameter= 0.8 arcsec $\sim$ 340 pc). The third column shows the average extinction values of the dusty ring (r$_1$ and r$_2$= 0.6 and 1.6 arcsec; dr= 1arcsec$\sim$430 pc). The three first rows display the E(B-V) values from the gas at blueshifted, central and redshifted velocities respectively. The four to the last rows display the intrinsic extinctions of the central and redshifted gas. The contribution of inflows to the extinction is taken into account in the rows with subscript \textit{INFLOW$\_$PERCENT}.
    }
    \label{tab:EBV_int_dusty_bubble_knotB}
\end{table*}

\begin{figure*}
    \centering
    \includegraphics[width=12cm]{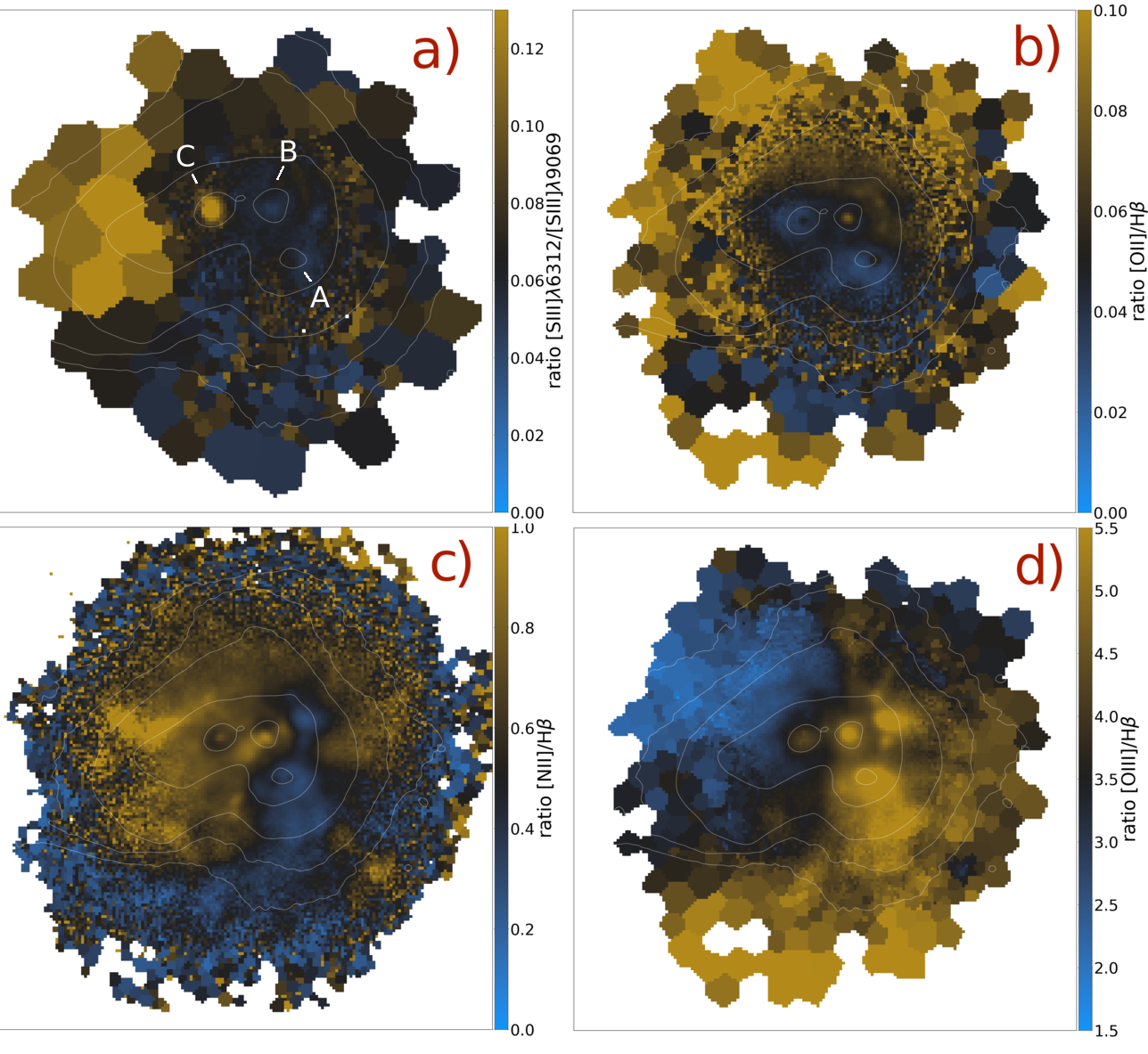}
    \caption{a) Line ratio of the [S III] auroral to the nebular line used to derive the temperature. This panel shows clearly the enhancement of the auroral line in the shocked region and in knot C. b), d) and c) are line ratios used to derive the O$^{+}$, O$^{2+}$ and N$^{+}$ abundances respectively. The \OIIHb\ ratio (b) shows the regions where the abundance of the relative low ionized gas, more precisely $\alpha$-elements is high. The \OIIIHb\ ratio (d) is used to disentangle the regions with high and low ionization. As shown in \citet{menacho2019}, the greatest fraction of Haro 11 are highly ionized due to the intense radiation released by the massive star clusters in the knots. In the shocked area, the gas is predominantly neutral or lowly ionized. The \NIIHb\ ratio (c) shows the regions enhanced by nitrogen ions that are produced in Wolf-Rayet stars or they are released later on in the final phase of intermediate-mass stars. Nitrogen is mainly produced in Knot B and C while there is a deficit in knot A. Because of the latter, there is a deficit of nitrogen in the highly ionized southern hemisphere.
    }
    \label{fig:ratios}
\end{figure*}

\begin{figure*}
    \centering
    \includegraphics[width=2\columnwidth]{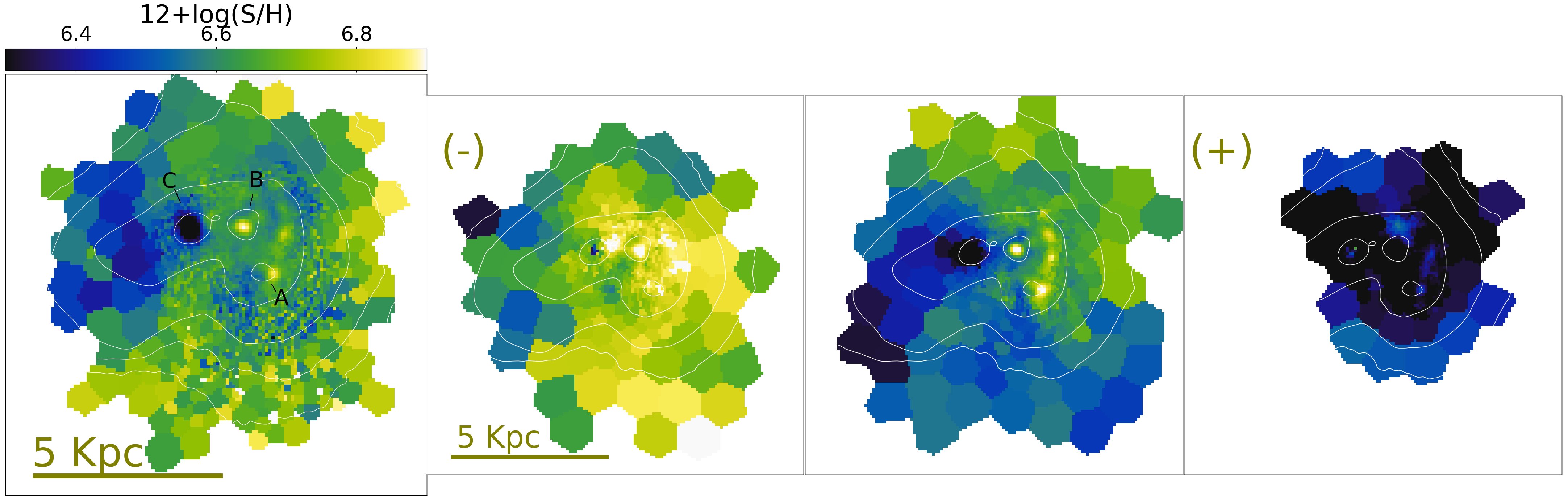} 
    \caption{Sulphur abundance with the same pattern than the oxygen abundance derived from the direct method. 
    }
    \label{fig:element_abundance_S}
\end{figure*}


\bsp	
\label{lastpage}
\end{document}